\documentclass{article}
\usepackage[utf8]{inputenc}

\usepackage{cite}
\usepackage{epsfig}
\usepackage[english]{babel}
\usepackage{a4wide}
\usepackage{amsmath}
\usepackage{amsthm}
\usepackage{amssymb}
\usepackage{units}
\usepackage{booktabs}
\usepackage{setspace}
\usepackage[a4paper,left=18mm,right=18mm, top=3.5cm, bottom=4cm]{geometry}
\usepackage{ifthen}
\usepackage{color}
\usepackage{framed}
\usepackage{bm}
\usepackage{nameref}
\usepackage{relsize}
\usepackage{ccaption}
\usepackage{hyperref}
\usepackage{fdsymbol}
\usepackage{appendix}

\usepackage{lineno}
  
\makeatletter
\renewcommand{\@biblabel}[1]{\quad#1.}
\makeatother

\newcommand{\KL}[2]{\mathcal{D}_\mathrm{KL}\bigl(#1\,\big\|\, #2 \bigr)}

\newcommand{\expect}[2][]{\left\langle\, \rule{0pt}{11.5pt} #2\, \right\rangle_{#1} }

\definecolor{gray}{gray}{0.2}



\newcommand{\prob}[1]{p\left({#1}\right)}
\newcommand{\cprob}[2]{p\left({#1}\,\middle\vert\,{#2}\right)}

\newcommand{\fun}[2]{{#1}\left( #2 \right)}
\newcommand{\cfun}[3]{\fun{#1}{ #2 \,\middle\vert\, #3 }}

\title{A synapse-centric account of the free energy principle}

\author{David Kappel and Christian Tetzlaff} 
\date{\today}

\begin{document}

\maketitle

\begin{abstract}
The free energy principle (FEP) is a mathematical framework that describes how biological systems self-organize and survive in their environment. This principle provides insights on multiple scales, from high-level behavioral and cognitive functions such as attention or foraging, down to the dynamics of specialized cortical microcircuits, suggesting that the FEP manifests on several levels of brain function. Here, we apply the FEP to one of the smallest functional units of the brain: single excitatory synaptic connections. By focusing on an experimentally well understood biological system we are able to derive learning rules from first principles while keeping assumptions minimal. This synapse-centric account of the FEP predicts that synapses interact with the soma of the post-synaptic neuron through stochastic synaptic releases to probe their behavior and use back-propagating action potentials as feedback to update the synaptic weights. The emergent learning rules are regulated triplet STDP rules that depend only on the timing of the pre- and post-synaptic spikes and the internal states of the synapse. The parameters of the learning rules are fully determined by the parameters of the post-synaptic neuron model, suggesting a close interplay between the synaptic and somatic compartment and making precise predictions about the synaptic dynamics. The synapse-level uncertainties automatically lead to representations of uncertainty on the network level that manifest in ambiguous situations. We show that the FEP learning rules can be applied to spiking neural networks for supervised and unsupervised learning and for a closed loop learning task where a behaving agent interacts with an environment.
\end{abstract}

\section{Introduction}
\label{sec:intro}

Synapses are inherently unreliable in transmitting their input to the post-synaptic neuron. For example, the probability of neurotransmitter release, a major source of noise in the post-synaptic potential (PSP) is typically around 0.5 \cite{katz1971quantal, oertner2002facilitation, jensen2019multiplex}, and can be as low as 0.2 in vivo \cite{borst2010low}, suggesting that up to 80\% of synaptic transmissions fail. This and other pre- and post-synaptic mechanisms result in a large trial-by-trial variability in the PSP \cite{rusakov2020noisy}.
Several authors have suggested that noisy synaptic transmission is a feature -- not a bug -- that enables the brain to reason about its own uncertainty \cite{maass2014noise, aitchison2014probabilistic, neftci2016stochastic, rusakov2020noisy, aitchison2021synaptic}, but a definite answer on the role of noise in synaptic transmission is still missing. Here, we show that synapses can exploit PSP noise to encode uncertainty about the somatic membrane potential of the post-synaptic neuron. More precisely, we show that synapses interact with the post-synaptic neuron by following the same principle of an organism that interacts with the world that surrounds it and PSP variability expresses the uncertainty of the synapse about its environment.

\begin{figure}[ht]
	\centering
	\includegraphics[width=\textwidth]{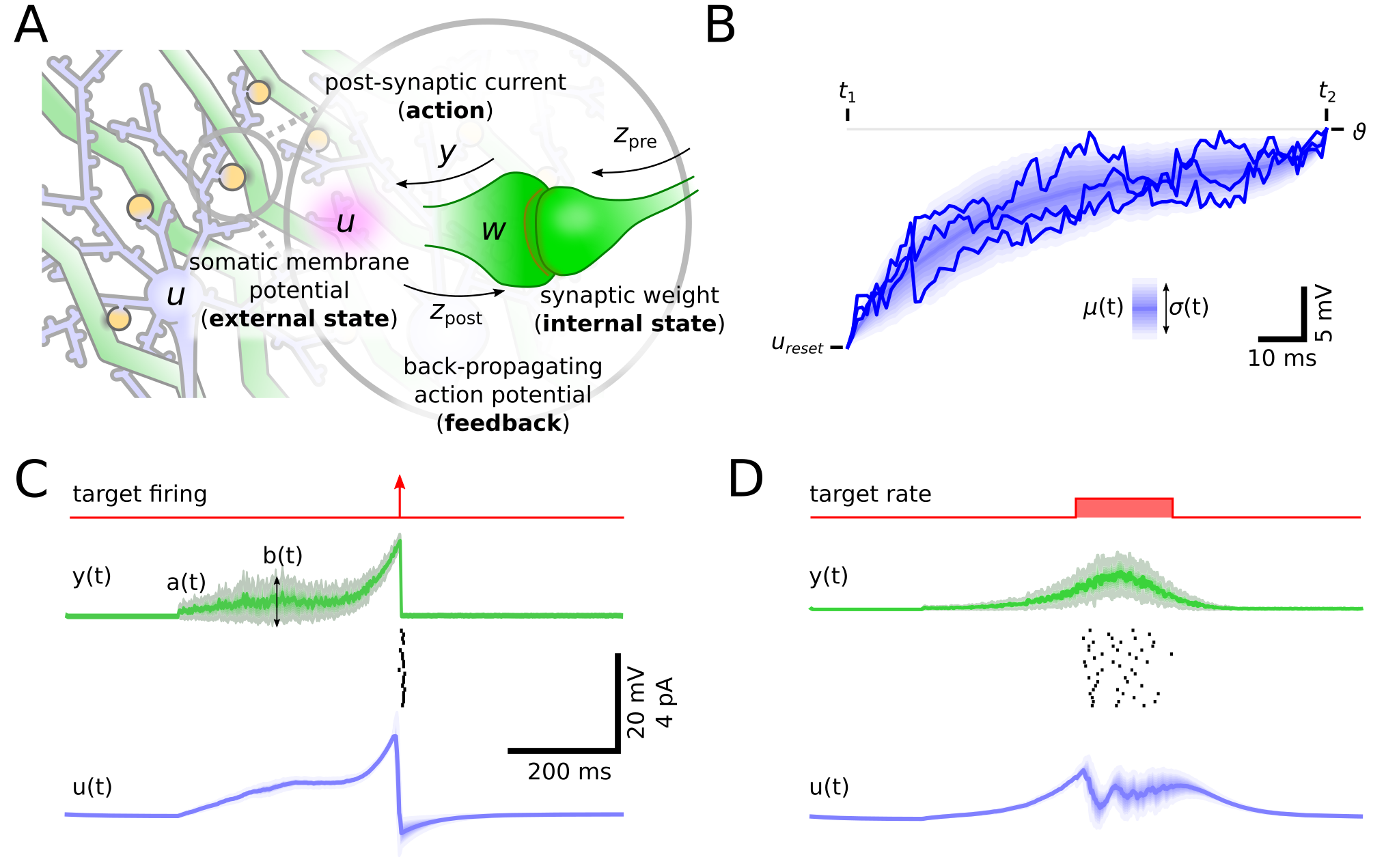}
	\caption{
	\textbf{The free energy principle for single synapses.}
		\textbf{A:} Illustration of the synapse model that interacts with its environment. Relevant variables are the post-synaptic current (action) the somatic membrane potential of the efferent neuron (external state), the back-propagating action potential (feedback), and the synaptic weight (internal state).
		\textbf{B:} Three individual trials and estimated probability density of the membrane potential over 868 trials of a post-synaptic spike interval of 100\,ms ($t_1$,$t_2$). Solid blue line shows the mean, variance indicated by shaded area. The membrane potential is constraint to the firing threshold $\vartheta$ at spike times and then reset to $u_{reset}$ immediately after every spike.
		\textbf{C:} Input current $y(t)$ (green) and membrane potential $u(t)$ (blue) to produce target spiking behavior indicated (red). Spiking behavior over 20 trials is shown.
		\textbf{D:} Same as in (C) but for a brief firing burst as target activity.
		\label{fig:bridge-model}
    }
\end{figure}

To establish this result we rely on a widely used model framework to describe biological systems that act in uncertain environments: The \emph{free energy principle} (FEP), which is based on the idea that biological systems instantiate an internal model of their environment that allows them to take actions to minimize surprise \cite{friston2010free}. A mathematical formulation of surprise can be closely related to the physical notion of free energy, from which the FEP inherits its name. The FEP is successful in explaining biological mechanisms on various spatial and temporal scales, e.g. dendritic self-organization \cite{kiebel2011free}, network-level learning mechanism \cite{isomura2018vitro}, human behavior \cite{ramstead2016cultural} and even evolutionary processes \cite{ramstead2018answering}. Here, we pursue a bottom-up approach that takes advantage of the excellent scaling abilities of the FEP. Moreover, by virtue of this principle the whole network automatically follows the FEP through an emergent effect of synapse-level FEP. 

The intuition behind our model is illustrated in Fig.~\ref{fig:model-illustration}A. Despite its apparent simplicity a synapse has all relevant components required by the FEP:  (1.) \emph{actions} in the form of post-synaptic current $y$ that enable the synapse to interact with (2.) \emph{the external states} of the environment, given by the somatic membrane potential $u$, (3.) \emph{feedback} in the form of the back-propagating post-synaptic action potential (spike) that allow the synapse to update the (4.) \emph{internal states} in the form of the synaptic weight $w$.

We analyze a learning rule that is directly derived from minimizing the free energy for learning in single synapses that interact with their environment. The FEP enables synapses to adapt their internal states to best predict future stimuli. The synapse does so by probing the environment using its noisy post-synaptic current and integrating the resulting feedback provided by the back-propagating action potential. Applied to a single neuron the emergent synaptic plasticity rule reproduces a number of experimentally observed mechanisms of LTP/LTD protocols. More precisely we show that the rule is well described by a regulated triplet STDP rule \cite{pfister2006triplets,gjorgjieva2011triplet}. Applied to the network level we show that the rule leads to self-organization and can be used to learn input-/output- dependencies of external stimuli.

Other than previous approaches (e.g. \cite{isomura2016linking}) that studied free energy minimization in the light of reward mechanisms like the dopaminergic system, we focus here on self-organization that emerges from only three local variables at the synapse: the pre-/post spike time and the current value of synaptic strength. We show that these variables play together in the FEP to enable an efficient learning machinery. Through stochastic synaptic currents synapses probe their environment and integrate the arriving feedback to update their internal state. Therefore, every stochastic release event can be seen as a small experiment, that is based on previous experience and the outcome of which shapes subsequent future activity. We show that this scheme gives rise to a viable learning model that can be scaled up to network-level tasks.

\section{Results}
\label{sec:results}

\subsection{Synaptic free energy model}

Here, we summarize the relevant steps to establish our model for free energy minimization on the synaptic level, a detailed and more formal derivation can be found in \nameref{sec:methods}. We start by defining the relevant components required by the FEP: (1.) \emph{the actions}, (2.) \emph{the external states}, (3.) the \emph{feedback} and \emph{internal states} (see \cite{Friston:08} and Fig.~\ref{fig:bridge-model}A for an illustration).
\begin{enumerate}
\item \emph{The actions}, that are utilized by the synapses to interact with the environment (the efferent neuron). In our model this is done through stochastic synaptic currents $y$, where the mean and variance of $y$ is governed by the synaptic strength $w$.
\item \emph{The external states}. From the perspective of a synapse the environment, it can immediately interact with, is the post-synaptic neuron. Here, we model the external states as the membrane potential $u(t)$ of a leaky integrate and fire (LIF) neuron with firing threshold $\vartheta$ and resting potential $u_0$.
\item \emph{The feedback}. In our model, a synapse only receives the back-propagating action potential of the post-synaptic neuron $z_{post}$ as feedback to be informed about the somatic membrane potential. Formally, the spike train $z_{post}$ is denoted by the set of firing times $t_{post}^{(n)}, t_{post}^{(n+1)}, \dots$ of the post-synaptic neuron. This feedback information about the external state $u(t)$ is used by the synapse to update the internal model of the environment.
\item \emph{The internal states} summarizes all relevant internal variables that determine the behavior of the synapse. Since we focus here on long term plasticity the internal state is given by the synaptic weight $w$. The internal states can be augmented with additional variables to also include other mechanisms, e.g. short term plasticity, but we neglect these here for the sake of simplicity.
\end{enumerate}

The FEP provides a generic approach to solve the \textit{internal state} $\rightarrow$ \textit{action}  $\rightarrow$ \textit{external state} $\rightarrow$ \textit{feedback} -loop in Fig.~\ref{fig:bridge-model}A, by minimizing the variational free energy $\mathcal{F}(z,w) = \text{surprise}(z) + \text{divergence}(z|w)$. The variational free energy $\mathcal{F}(z,w)$ measures the surprise caused by the feedback $z$ and the divergence between the internal model and the estimated external state based on the internal state $w$. To this end the synapse maintains an internal probabilistic model and uses stochastic synaptic currents to test this model against perception and in turn updates the internal state $w$. In our model synaptic currents $y$ are drawn from a Gaussian distribution parametrized by the synaptic weight $w$. Whenever a pre-synaptic input spike arrives at time $t$ synaptic currents are generated according to $y(t) \sim \cfun{\mathcal{N}}{y}{r_0\,w,s_0\,w}$, where $r_0$ and $s_0$ are scaling constants for the mean and variance, respectively. These stochastic synaptic currents capture the combined effect of pre- and post-synaptic noise sources, such as stochastic synaptic release.

The FEP explains the behavior of the synapse as the solution to a planning problem, where every synapse strives to producing synaptic currents $y$ that are consistent with the back-propagating action potentials. This means that $y$ should best match the evolution of $u$ which leads to the spiking activity $z$, i.e. the \textit{action}  $\rightarrow$ \textit{external state} $\rightarrow$ \textit{feedback} dependency. But the synapse does not have access to the true value of the somatic membrane potential $u$ and therefore it has to be inferred from the sparse information that is contained in the feedback $z$. This is captured in a model $\cprob{u}{z}$ that expresses the probability density over membrane potential trajectories for given spike trains $z$. The back-propagating action potential $z_{post}$ only conveys the information that the post-synaptic membrane potential has just reached the firing threshold $u(t)=\vartheta$. This means that at any moment in time $z_{post}$ provides a single bit of information about $u(t)$, which encodes whether the firing threshold $\vartheta$ has been crossed at time $t$. Otherwise the membrane potential $u(t)$ evolves according to some unobserved dynamics which includes all synaptic input arriving at the somatic compartment, which leads to high trial-to-trial variability.

To illustrate the information that can be accessed by a synapse we simulated a single neuron that received random input generated by noisy synaptic currents with constant mean and variance in Fig.~\ref{fig:bridge-model}B. This input randomly drives the membrane to reach the firing threshold with different inter spike time intervals ($\Delta t = t_2 - t_1$). To analyze the trial-by-trial variability of the membrane voltage we show individual traces with a fixed $\Delta t$=100\,ms.  By taking averages over many traces we can recover the statistical properties of the membrane potential evolution (mean $\mu(t)$ and variance $\sigma^2(t)$ over 868 trials indicated by blue shaded area). This setup provides us with an empirical estimate of $\cprob{u}{z}$ for a given $\Delta t$. In Appendix \ref{sec:apx-prediction-density} we show that $\cprob{u}{z}$ can be expressed analytically for arbitrary $\Delta t$. All these solutions have in common that the uncertainty ($\sigma^2(t)$) is minimal close to the firing times and gradually increases reaching its maximum at around $\frac{\Delta t}{2}$.

The FEP is a model-based approach that maintains $\cprob{u}{z}$ as an internal representation of the dynamics of the external state. The FEP also explicitly expresses the uncertainty about the state of the environment, which can be determined by time-varying mean and variance functions, $\mu(t)$ and $\sigma^2(t)$, respectively. This internal model allows us to answer queries about the external world, i.e. what are synaptic currents $y$ that most likely lead to a spiking behavior $z$. Here we use a stochastic leaky integrate and fire (LIF) neuron with resting potential $u_0$ and membrane time constant $\tau_m$ to describe the dynamics of the internal model.
In \nameref{sec:methods} we show that for any mean and variance function $\mu(t)$ and $\sigma^2(t)$ of the membrane potential $u(t)$, we can infer a distribution over synaptic currents $y(t)$ that will lead to its realization. This is achieved by choosing $y(t) \sim \cfun{\mathcal{N}}{y(t)}{a(t),b(t)}$, with $a(t) = \mu(t)' + \frac{1}{\tau_m} (\mu(t)-u_0)$ and $b(t) = \left( \sigma^2(t) \right)' + \frac{2}{\tau_m} \sigma^2(t)$, where $\mu(t)'$ and $\left( \sigma^2(t) \right)'$ denote time derivative. Using this, arbitrary membrane potential dynamics and firing patterns can be realized.

Fig.~\ref{fig:bridge-model}C and D show two examples.
In Fig.~\ref{fig:bridge-model}C we used a current $y(t)$ that leads to a similar behavior to the trial-averaged dynamics in Fig.~\ref{fig:bridge-model}B. The membrane potential reaches the threshold $\vartheta$ at a predetermined firing time. The probabilistic model does not only allow us to define fixed firing times but also to create target distributions of firing times as shown in Fig.~\ref{fig:bridge-model}D, given here by a brief burst of neural firing. We used this target to infer distributions over synaptic current and membrane potential. Trial averages over 20 runs are shown.

\subsection{Synaptic plasticity as free energy minimization}

\begin{figure}[ht!]
	\centering
	\includegraphics[width=\textwidth]{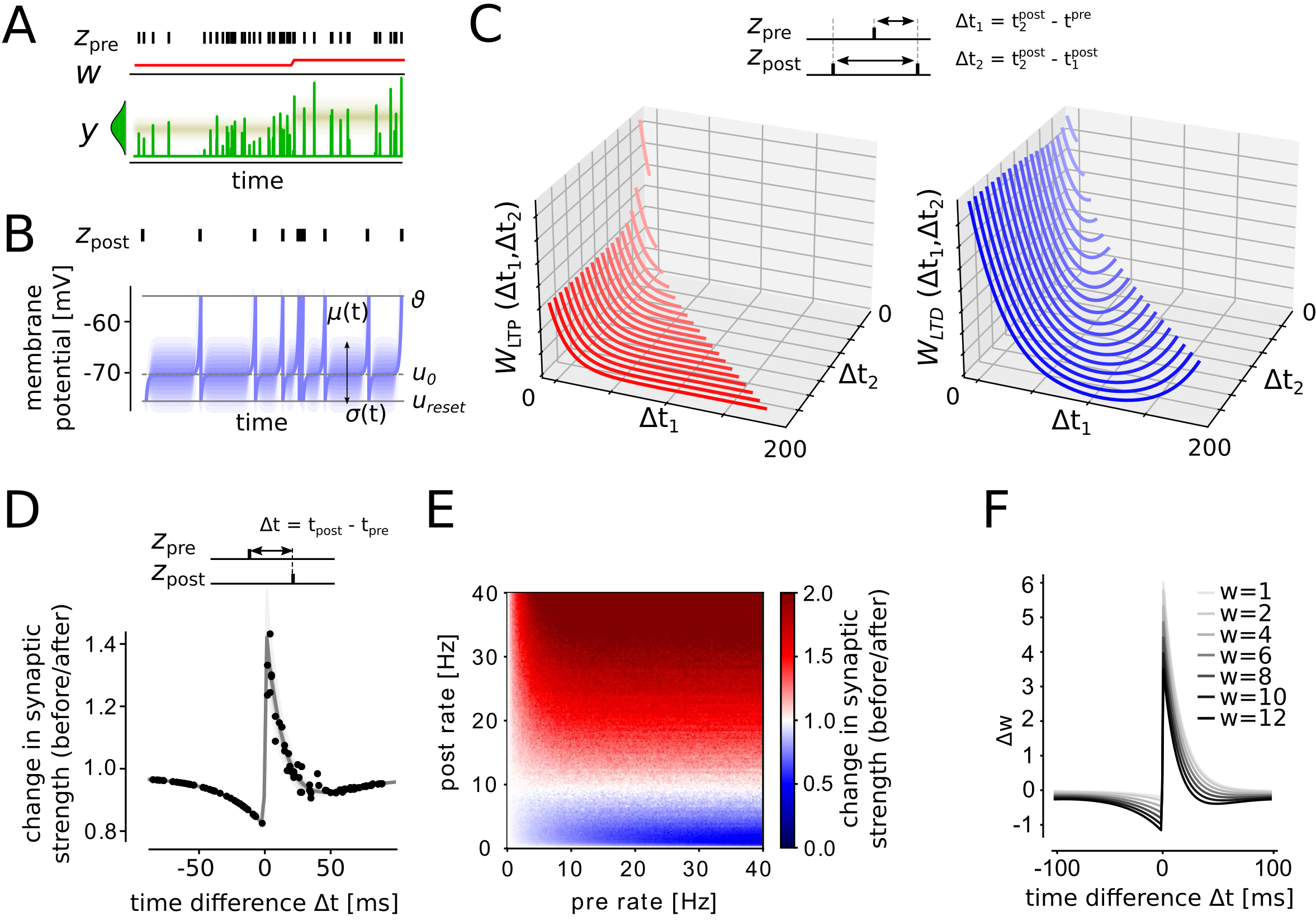}
	\caption{
	\textbf{Regulated triplet STDP enables synapse-level FEP.}
		\textbf{A:} Synapses cause stochastic post-synaptic currents in response to pre-synaptic input spikes.
		\textbf{B:} Probability density of the membrane potential according to the stochastic process $(\mu(t), \sigma(t))$. Solid blue line shows the mean $\mu(t)$, $\sigma(t)$ indicated by shaded area. The membrane potential is constraint to the firing threshold $\vartheta$ at spike times and then reset to $u_{reset}$ immediately after every spike.
		\textbf{C:} The triplet STDP windows $W_{LTP}$ and $W_{LTD}$ that emerge from the FEP learning model.
		\textbf{D:} Mean synaptic weight changes (gray line) and individual trials (black dots) for an STDP pairing protocol.
		\textbf{E:} Synaptic weight changes as a function of pre- and post- rate.
		\textbf{F:} Weight dependence of the FEP learning rule.
		\label{fig:model-illustration}
    }
\end{figure}

In the previous section we have shown that the dynamics of $u(t)$ can be expressed explicitly by a stochastic process $(\mu(t), \sigma(t))$, which denotes the solution of the LIF dynamics constraint to the (only) known values $u(t_{post}^{(n)})=\vartheta$ at the post-synaptic firing times $t_{post}^{(n)}$. Furthermore, the FEP allows us to provide solutions for the synaptic current $y$ that leads to an observed firing behavior $z$.

Here we show how the FEP can be used to derive learning rules for the synaptic weights $w$. Whenever an action potential is triggered by pre-synaptic inputs the synapse produces a brief current pulse according to its learned internal model represented by the synaptic weight $w$. As back-propagating action potentials invade the synapse the synaptic weight is updated to more closely match the desired firing activity. This allows us to analytically express the free energy and derive a learning rule that minimize $\mathcal{F}(z,w)$ with respect to $w$ (see \nameref{sec:methods} for a detailed derivation). Since the evolution of the membrane statistics $\mu(t)$ and $\sigma^2(t)$ only depend on the back-propagating action potentials their dynamics can be fully determined by the relative firing times (see Fig.~\ref{fig:model-illustration}A,B for an illustration).
The synaptic weight updates therefore have the form
\begin{equation}
\Delta\,w \;=\;
        \fun{W_{LTP}}{\Delta t_1, \Delta t_2} \;-\; 
        \left( \frac{1}{2} + w \right) \fun{W_{LTD}}{\Delta t_1, \Delta t_2}
        \;+\; \frac{1}{2\,w} \;.
        \label{eqn:dw}
\end{equation}
where $W_{LTP}(\Delta t_1, \Delta t_2)$ and  $W_{LTD}(\Delta t_1, \Delta t_2)$ are triplet STDP learning windows that depend only on time differences $\Delta t_1 = t_2^{post} - t^{pre}$ and $\Delta t_2 = t_2^{post} - t_1^{post}$ of neighboring pre- and post-synaptic spikes, and $w$ denotes the current value of the synaptic efficacy.

The functional form of the triplet STDP windows is shown in Fig.~\ref{fig:model-illustration}C. $W_{LTP}$ has a potentiating effect which is maximal close to $t_2^{post}$. This is a manifestation of Hebbian-type learning where close correlations of pre- before post- firing leads to potentiation. $W_{LTD}$ has a depressing influence on the synaptic weight. It peaks on both sides when close to a post-synaptic spike and is at its minimum around $\frac{\Delta t_2}{2}$ where the uncertainty about $u(t)$ is maximal. Both STDP windows show also a strong rate dependence ($\Delta t_2$) as higher firing rates result in overall less uncertainty about $u(t)$. 
In addition the learning rule \eqref{eqn:dw} shows a weight dependence that regulates the synaptic strength. The two triplet STDP windows depend on the time differences $\Delta t_1$ and $\Delta t_2$ in a nonlinear manner \cite{pfister2006beyond}. Fig.~\ref{fig:model-illustration}C shows the shape of the STDP windows $W_{LTP}(\Delta t_1, \Delta t_2)$ and  $W_{LTD}(\Delta t_1, \Delta t_2)$.

\subsection{Synapse-level FEP is compatible with STDP and Calcium-based plasticity}

In this section we identify the most salient properties of the FEP learning rule Eq.~\eqref{eqn:dw}.
To test our learning rule we put it in a synaptic environment. First, we applied an STDP pairing protocol where single pre-/post spike pairs with different time lags $\Delta t$ were presented to a model synapse and synaptic changes were measured with respect to $\Delta t$ \cite{pfister2006beyond}. The results after applying 10 pre-/post pairs are shown in (Fig.~\ref{fig:model-illustration}D). The learning window closely matches experimentally measured STDP windows \cite{dan2004spike, caporale2008spike}.

In Fig.~\ref{fig:model-illustration}E we further study the rate dependence of our learning rule. Random pre- and post-synaptic Poisson spike trains where generated with different rates. The resulting synaptic weight changes after learning for 10 seconds were measured. For low pre- or post-synaptic rates synaptic weight changes were zero. Moderate post-synaptic rates lead to LTD, whereas high post-synaptic rates manifested in LTP. This effect is consistent with previous models of calcium-based plasticity \cite{graupner2012calcium}.

In Fig.~\ref{fig:model-illustration}F we analyze the weight dependence of the learning rule. The learning rule Eq.~\eqref{eqn:dw} automatically regulates the synaptic weight to not grow out of bounds. To show this we applied STDP protocols for synapses with different initial synaptic weights. Small synaptic weights ($w$=1\,pA) lead to learning windows that are positive for all lags $\Delta t$ (LTP only). Large synaptic weights $w$=12\,pA lead to pronounced LTD behavior.

In summary these results show that the FEP learning rule show features of classical Hebbian learning, spike-timing-dependent plasticity (STDP) and rate dependent learning rules. More precisely, our learning rule can be best described by an STDP triplet rule that depends on the pre-synaptic and the two neighboring post-synaptic spike times (post-pre-post triplet STDP rule \cite{pfister2006beyond,pfister2006triplets,gjorgjieva2011triplet}). Our results suggests that a pre-synaptic spike that arrives briefly before a post-synaptic action potential should cause long term potentiation (LTP), whereas spikes arriving shortly after should lead to depression (LTD), much like in many other STDP or Hebbian learning rules that have been suggested. In addition our learning rule amplifies post-synaptic high frequency events (see Fig.~\ref{fig:model-illustration}E) and it shows a strong weight dependence (Fig.~\ref{fig:model-illustration}F). The strength of synaptic depression increases with the efficacy of the synapse $w$. This gives rise to a homeostatic effect that prevents synapses from growing out of bounds. 

\subsection{Synapse-level probability matching of firing times}

\begin{figure}
	\centering
	\includegraphics[width=1.0\textwidth]{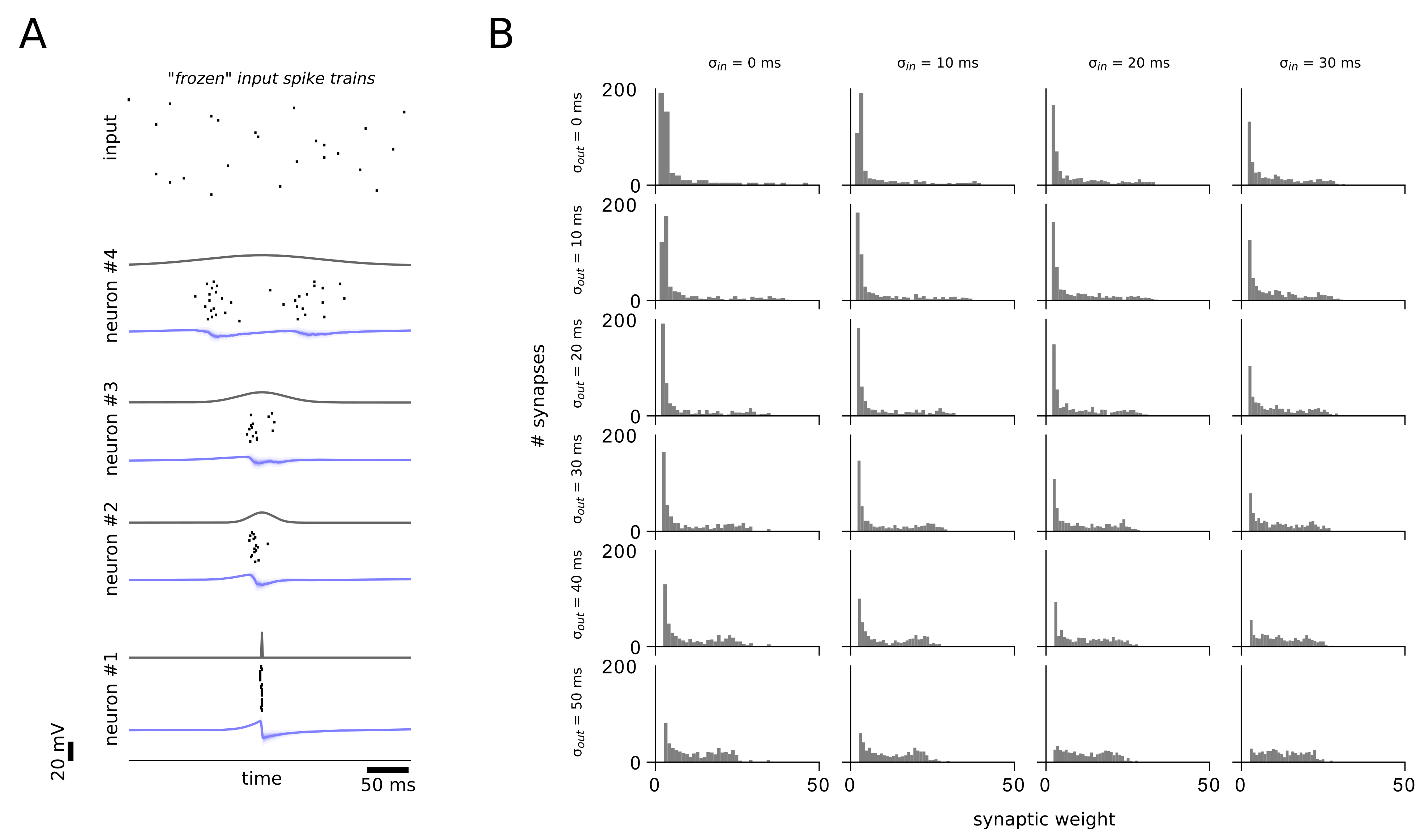}
	\caption{\textbf{Synapse-level probability matching.} \textbf{A:} A single neuron was presented with a frozen input spike train over 200 input neurons (top). The post-synaptic neuron was brought to fire according to a probability distribution given by a single pulse (neuron~\#1) or a Gaussian distribution with different deviations $\sigma_{out}$ (neuron~\#2-\#4). Target spiking behavior after learning (solid gray line), individual output spikes over 20 runs (black dots) and trial-averaged membrane potentials (blue) are shown.
	\textbf{B:} Histograms over emergent synaptic weights after learning output spikes with different input ($\sigma_{in}$) and output ($\sigma_{out}$) spike time deviations.
	}
 \label{fig:probability-matching}
\end{figure}
A prevailing feature of the FEP is that agents that follow this principle acquire an implicit probabilistic representation of their environment, that captures its typical behavior and its variability. After the model of the environment has been acquired it can be used to reproduce state trajectories that match the learned probabilistic model.

To demonstrate this mechanism for our synaptic FEP model, we consider here a single neuron that receives input from afferent neurons that fire according to a certain random input spike train, where every neuron emits a sparse spike train in a time window of 300\,ms. This input spike train is repeatedly presented to the neuron and the neuron is driven externally to fire according to a defined target spike distribution (see Fig.~\ref{fig:probability-matching}). To demonstrate the learning capabilities of the proposed model we applied different target distributions with different variances. Neuron~\#1 in Fig.~\ref{fig:probability-matching}A was brought to fire a single spike at 150\,ms after input onset. Neurons~\#2-\#4 were brought to fire according to a Gaussian distribution with mean at 150\,ms and different spread $\sigma_{out}$ of 20-100\,ms. After learning the neuron was able to reproduce the firing distribution (spike trains in Fig.~\ref{fig:probability-matching} show spiking behavior of 20 individual trials). The membrane potential reflects the dynamics of the target spike trains. During the phase of stochastic firing we observe a high trial-to-trial variability in the dynamics of the membrane potential (Fig.~\ref{fig:probability-matching}A). Note that the input and the LIF neuron model are deterministic here, so the required trial-by-trial variability is produced exclusively by the synapses. Hence, synapses have learned to utilize their intrinsic variability to drive the deterministic neuron to fire according to a defined probability distribution.

In Fig.~\ref{fig:probability-matching}B we further analyze the learning behavior for synapse-level probability matching. Here we used in addition to different output divergence $\sigma_{out}$ also stochastic input spike times that were drawn from a Gaussian with divergence $\sigma_{in}$. Weight histograms are shown over all synapses of 5 individually trained neurons for each ($\sigma_{in}$, $\sigma_{out}$) pair. The weight histograms reflect the task that was to be learned. If high output precision is required (e.g. $\sigma_{out} = 0$) few very strong synapses are formed and the overall spike distribution has a heavy-tailed shape. With higher input and output variability the weight histogram approaches a uniform distribution.

\subsection{Network-level learning using the synaptic FEP}

\begin{figure}
	\centering
	\includegraphics[width=\textwidth]{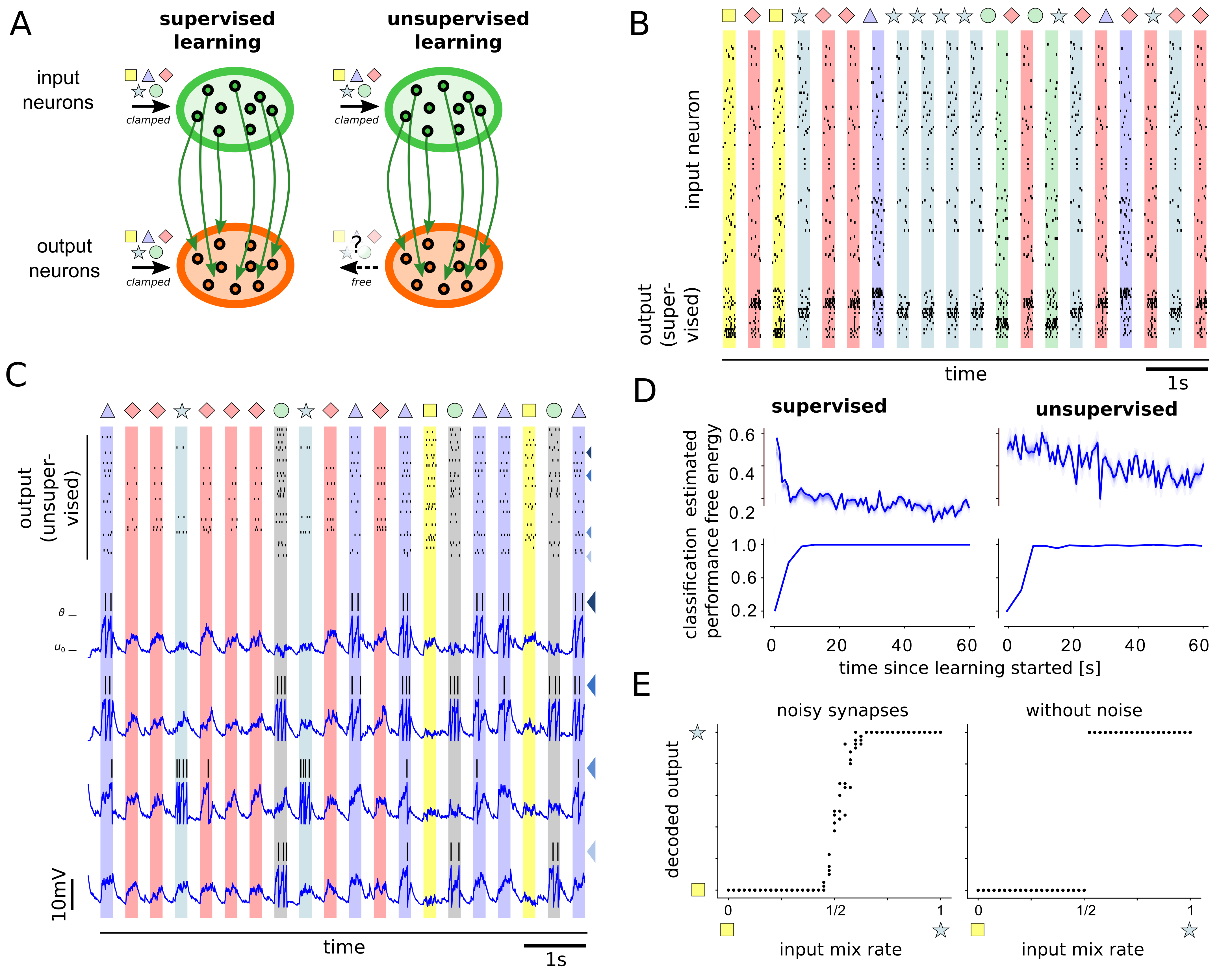}
	\caption{\textbf{The FEP for supervised and unsupervised learning.}
	\textbf{A:} Illustration of the network structure. Five independent spike patterns ($\square$,$\largewhitestar$,$\triangle$,$\meddiamond$,$\largecircle$) are presented to the network by clamping the input neurons. Output neurons are either clamped to pattern-specific activity during learning (supervised) or allowed to run freely (unsupervised). 
	\textbf{B:} Learning result using the synapse-level FEP rule for the supervised scenario. Typical spiking activity of the network after learning for 60\,s. Black ticks show output spike times.
	\textbf{C:} Output activity after learning for the unsupervised scenario. Traces of membrane potentials are shown for selected output neurons (matching color-coded arrows indicate neuron identities).
	\textbf{D:}
	Classification Performance and estimated free energy for supervised and unsupervised learning scenario. The estimated free energy per synapse decreases with learning time. Classification performance plateaus at near optimal value after about 20\,s of learning time for both supervised and unsupervised scenario.
	\textbf{E:}
	Spike patterns of two input symbols ($\square$, $\largewhitestar$) where mixed with different mixing rates. Uncertainty is reflected in output decoding (left) if inputs are ambiguous (around mixing rate of 1/2). If synapse noise is disabled uncertainty is not represented in the output (right).
	}
 \label{fig:pattern-classification}
\end{figure}
The FEP lends itself very well to supervised and unsupervised learning. To demonstrate this for our synapse-based FEP model we consider a pattern classification task. The network architecture is shown in Fig.~\ref{fig:pattern-classification}A. The network consists of input neurons that project to a set of output neurons. We generated five spike patterns of 200\,ms duration (denoted in Fig.~\ref{fig:pattern-classification} by $\square$,$\largewhitestar$,$\triangle$,$\meddiamond$ and $\largecircle$) which were used to control the activity of the input neurons. Pattern presentations were interleaved with phases of 200~ms of zero spiking on all input channels. In the supervised scenario for every output neuron one of the five patterns was selected as preferred stimulus. During training the activity of the output neurons was clamped to fire during the presentation of the preferred stimulus pattern. In the unsupervised case output neurons were simply allowed to run freely according to their intrinsic dynamics. In the unsupervised case a rate adaptation was used to prevent the output neurons from becoming silent (see \nameref{sec:methods} for details). The FEP learning rule was active for all synapses between input and output neurons in both scenarios.

Fig.~\ref{fig:pattern-classification}B shows the typical network activity after learning for 60\,s for the supervised scenario. The output neurons reliably responded to their preferred pattern. The output neurons had also learned a sparse representation of the input patterns in the unsupervised case (Fig.~\ref{fig:pattern-classification}C). Most neurons (46/50) were active during exactly one of the input patterns (e.g. the $\triangle$-selective neuron in the top row of Fig.~\ref{fig:pattern-classification}C). The remaining neurons showed mixed selectivity and thus got activated by multiple stimulus patterns (see bottom rows of Fig.~\ref{fig:pattern-classification}C).

Fig.~\ref{fig:pattern-classification}D shows the evolution of the estimated mean free energy per synapse and the classification performance throughout learning. The free energy decreased on average throughout the learning process in both scenarios. After learning for 60\,s the pattern identity could be recovered by a linear readout with 100\% and 98.8\% reliability for the supervised and unsupervised case, respectively (see Fig.~\ref{fig:pattern-classification}D bottom). These results demonstrate that the FEP learning rule can be applied to supervised learning and also leads to self-organization of meaningful representations in an unsupervised learning scenario.

The uncertainty encoded on the synapse level can also be read out from the neural activity. To demonstrate this we created ambiguous patterns by mixing the spikes of two patterns ($\square$ and $\largewhitestar$) with different mixing rates. Mixing rates of 0 (1) corresponds to a pattern that is identical to $\square$ ($\largewhitestar$) and intermediate values gave results of different ambiguity. High levels of ambiguity were also encoded in the neural output (Fig.~\ref{fig:pattern-classification}E, left). Noisy synapses were necessary for encoding of uncertainty. To test this we trained a second network with noise turned off (Fig.~\ref{fig:pattern-classification}E, right). In this case the network uncertainty could not be decoded from the output activity, the decision between $\square$ and $\largewhitestar$ flipped around the maximum ambiguity at 1/2.

\subsection{Behavioral-level learning using the synapse-level FEP in a closed loop}

\begin{figure}
	\centering
	\includegraphics[width=\textwidth]{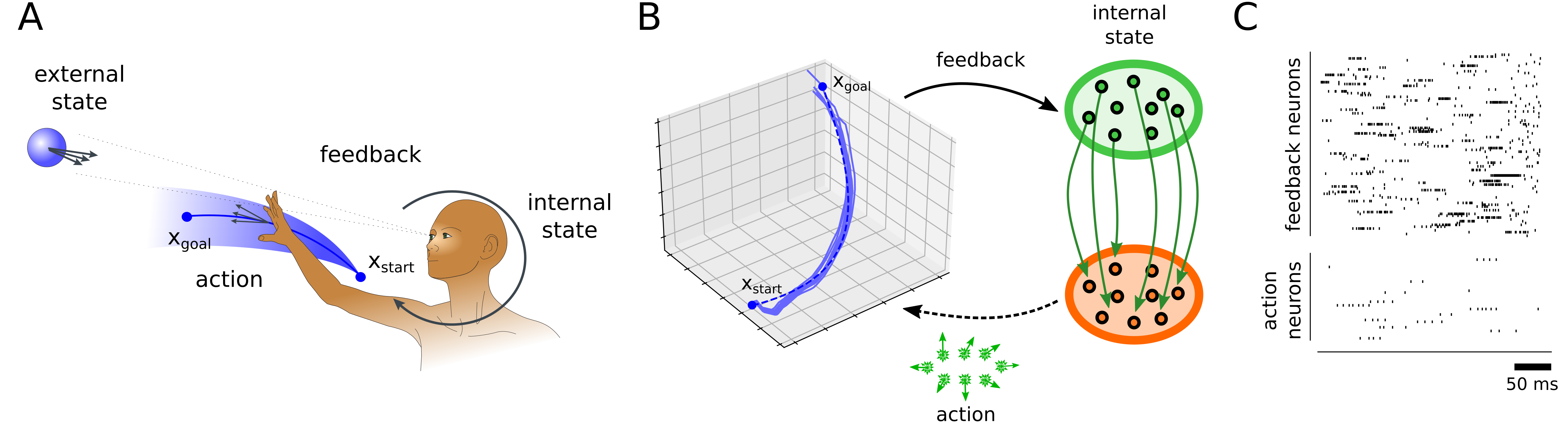}
	\caption{\textbf{The FEP for supervised and unsupervised learning.}
	\textbf{A:} Illustration of the behavior level FEP for an agent that interacts with a dynamic environment (adapted from \cite{faisal2008noise}). 
	\textbf{B:} A spiking neural network interacting with an environment using synapse-level FEP to learn a control policy. The activity of action neurons controls the movement of an agent in a 3-dimensional environment. Feedback about the position of the agent is provided through feedback neurons. The policy to navigate the agent is learned through synapse-level FEP between feedback and action neurons. Typical movement trajectories generated by the network are shown (blue).
	\textbf{C:}
	A typical spike train generated by the network after learning for 50\,s. The network is here allowed to freely interact with the environment after learning.
	}
 \label{fig:closed-loop}
\end{figure}
To further investigate the network effects of synapse level FEP, we implemented a closed loop setup where a spiking neural network controls a behaving agent. Many previous models have focused on how the FEP enables a behaving agent to interact with a dynamic environment \cite{friston2010free}. We provide here a proof-of-concept study how synapse-level FEP can be used as a building block to learn to interact with an environment on the behavior level. 

The behavioral level setup is illustrated in Figure~\ref{fig:closed-loop}A. The task here is to reach a fixed goal position $x_{goal}$ starting from $x_{start}$ in a 3-dimensional task space. The network that was used to learn this task is shown in Fig.~\ref{fig:closed-loop}B. It consists of a set of input neurons that receive encoded representations of the agents current position. A set of action neurons encodes preferred directions that are applied to update the agents position. The weights between feedback and action neurons were trained suing the synaptic FEP rule Eq.~\eqref{eqn:dw}. During training actions are given externally to provide a supervisor signal. Four typical trajectories after training for 50 seconds are shown (40 repetitions of the target trajectory where performed previously during training). Fig.~\ref{fig:closed-loop}C shows typical network activity of after training. The network has learned internal representations to reliably control the agent in a closed loop setting.

\section{Discussion}
\label{sec:discussion}

\subsection{Previous spiking network models and experimental evidence for the FEP and predictive coding}

The FEP and the much related theory of predictive coding have been very successful in explaining animal behavior and brain function \cite{rao1999predictive,friston2005theory,friston2010free,chalk2018toward}. On the neuron and network level 
previous models utilized the FEP to derive learning rules for reward-based learning and models of the dopaminergic system \cite{friston2014anatomy,isomura2016linking}.
In \cite{urbanczik2014learning} a model for dendritic prediction of somatic spiking was proposed. This model utilizes a two-compartment neuron model and learning depends on local membrane potential at the dendritic compartment which is updated to match the spiking behavior at the soma. In contrast to our model uncertainty about the membrane potential is not represented. The variational Bayesian inference method, which is at the core of the FEP, has been used to learn auto-encoder network dynamics in spiking neural networks \cite{deneve2008bayesian,BreaETAL:13,RezendeGerstner:14,RezendeETAL:11,RezendeETAL:14}.
There is also a close relationship between the FEP and the information theoretic measure of Shannon entropy. Previous models have demonstrated that spiking neurons can learn to minimize the loss of relevant information transmitted in the output spike train
\cite{buesing2008simplified,buesing2010spiking,linsker1988self}.

A number of previous studies have approached the problem of deriving learning rules from the free energy principle and other information-theoretic measures. \cite{isomura2016linking} used the FEP to derive synaptic weight updates with third factor modulation using dopamine-like signals. In \cite{toyoizumi2005generalized} it was shown that a variant of the Bienenstock–Cooper–Munro (BCM) rule for synaptic plasticity maximizes the mutual information between pre- and postsynaptic spike trains at single synapses. This result was generalized to show that a similar rule \cite{buesing2008simplified, buesing2010spiking} can perform information bottleneck optimization and principal component analysis in feed-forward spiking networks. These results are a special case of the general free energy minimization framework \cite{feinstein1986relating,tishby2000information,friston2012attractors}.

Direct experimental evidence for the FEP acting in cultured neurons was provided by 
\cite{isomura2015cultured} where it was discovered that neurons could learn to represent particular sources while filtering out other signals. Furthermore in \cite{isomura2018vitro} a Bayes-optimal encoding model was formulated and shown that these idealised responses could account for observed electrophysiological responses in vitro. Finally, evidence for predictive coding is abundantly available in in-vivo recordings of neural activity and brain anatomy \cite{BastosETAL:12,kanai2015cerebral, barascud2016brain,driscoll2017dynamic,kostadinov2019predictive}.

While plasticity of synaptic transmission probability has been documented \cite{markram1996redistribution, yang2013presynaptic, monday2018long} we focus here on a model where only the synaptic efficacy is plastic. The FEP suggests that all parameters of a biological system should evolve to minimize the free energy. In future work we will explore the role of plastic synaptic transmission probability to accurately learn complex spiking behavior.

\subsection{STDP and triplet STDP}

STDP is widely considered to provide a biological basis for the Hebbian postulate of correlation-based learning in the brain \cite{caporale2008spike}. Therefore STDP learning is often employed in theoretical models of synaptic plasticity. \cite{NesslerETAL:08,NesslerETAL:13,KappelETAL:14} demonstrated Bayesian learning capabilities of STDP in a cortical microcircuit motive. \cite{pecevski2014nevesim, pecevski2016learning} have demonstrated that STDP learning rules can learn arbitrary statistical dependencies between spike trains. \cite{pfister2006triplets} examined and formalized triplet STDP rules which considers sets of three spikes and compared them to classical pair-based STDP learning rules. They showed that triplet rules provide an excellent fit of experimental data from visual cortical slices as well as from hippocampal cultures. In \cite{gjorgjieva2011triplet} triplet rules were further analyzed and were found to be selective to higher-order spatio-temporal correlations, which exist in natural stimuli and have been measured in the brain.
\cite{clopath2010voltage,clopath2010connectivity} unifying STDP and voltage-dependent learning rules into a single model.

\subsection{Applications of the FEP in machine learning}

The FEP and predictive coding has also strongly influenced machine learning research. Most prominently in the literature on variational inference and auto-encoders (see e.g.  \cite{MnihGregor:14} for a recap). These models most often follow a top-down approach and are trained internally by the error back-propagation (Backprop) algorithm. However, more recently it was shown that the FEP may also provide an interesting alternative to Backprop. In \cite{whittington2017approximation} it was demonstrated that a special case of the FEP emulates the synaptic weight updates of Backprop with Hebbian-style learning rules. This work was recently generalized to emulate Backprop in arbitrary deep learning networks \cite{millidge2020predictive}. Following this line of research could provide a definite answer on how the brain manages to achieve its remarkable performance at a minimum   communication overhead.

Another important property of the learning algorithm is that synaptic updates only depend on the timing of pre- and post-synaptic spikes. 
The model is therefore very well suited for event-based neural simulation  \cite{pecevski2014nevesim,peyser2017nest}. Since in most applications the neural firing rate is quite low (typically in the range 0-5~Hz per neuron) the required processing power per synapse is also quite low. This property makes the model also appealing for new brain-inspired hardware \cite{mayr2019spinnaker,davies2018loihi}.

\subsection{Conclusion}

In summary, we have presented a synapse-centric account of the FEP that views synapses as agents that interact with their post-synaptic neuron much like an organism interacts with its environment. Using this principle we derive a learning rule based on very few assumptions. This learning rule matches experimentally observed synaptic mechanisms at a high level of detail. Our results complement previous applications of the FEP on the system and network level \cite{friston2010free,isomura2016linking} and demonstrates that manifestations of the FEP can be identified even on the smallest scales of brain function. In contrast to this prior work our model synapses use only local information and yields triplet STDP dynamics which can be directly tested against experiments. The emergent learning algorithm is fully event-based, i.e. computation only takes place when pre- and post-synaptic spikes arrive at the synapses. The model is therefore very well suited for event-based neural simulation  and brain-inspired hardware.

\section{Methods}
\label{sec:methods}

\subsection{Neuron model}
\label{sec:neuron}

We consider the leaky integrate and fire (LIF) neuron model \cite{gerstner2014neuronal}. The membrane potential $u(t)$ at the soma at time $t$ follows the dynamics
\begin{equation}
\tau_m \frac{d\,u}{d\,t}  \;=\; - \left(u(t) - u_0 \right) \;+\; R\,y(t)\;,
\label{eq:membrane-potential}
\end{equation}
where $\tau_m$ is the membrane time constant, $u_0$ is the resting potential, $R$ the membrane resistance and $y(t)$ is the total external input current into the neuron at time $t$ and denotes the summed effect of afferent synaptic inputs. 
When the membrane potential reaches the threshold $\vartheta$, the neuron emits an action potential, such that the spike times $t_f$ are defined as the time points for which the criterion
\begin{equation}
t_f\;: \quad u\left(t_f\right) = \vartheta\;,
\end{equation}
applies\cite{gerstner2014neuronal}. Immediately after each spike the membrane potential is reset to the reset potential $u_{reset}$.

\subsection{Synapse model and Learning rule}
\label{sec:synapse}

We use a stochastic synapse model of input-dependent current, where the variability rate is proportional to the synaptic efficacy $w$ \cite{yang2013stochastic}. The post-synaptic input current $y(t)$ is given by
\begin{equation}
    y(t) \;=\; z_{pre}(t)\left( w\,r_0 + \sqrt{w\,s_0} \,\epsilon(t) \right) \;,
    \label{eq:synaptic-current}
\end{equation}
where $w \geq 0$ is the synaptic efficacy, $z_{pre}$ is a spike train given by Dirac delta pulses centered at pre-synaptic spike times, and $\epsilon(t)$ is a source of independent unit variance zero mean Gaussian noise. The constant $r_0$ and $s_0$ scale the mean and variance of the synaptic current. We used $r_0=\frac{1}{2}$ and  $s_0 = r_0\,(1-r_0)$ if not stated otherwise in accordance with previous models. The synapse model \eqref{eq:synaptic-current} is a simplified Gaussian approximation to previous models of stochastic synaptic conductance that assumed a Binomial distribution over effective post-synaptic currents \cite{gontier2020identifiability, katz1971quantal}. The parameter $r_0$ can therefore be linked to the pre-synaptic release probability but captures here the combined effect of synaptic transmission noise. The Gaussian approximation emerges in the limit of a large number of synaptic release quanta and was used here to simplify the derivations.

The synaptic weights were updated using the learning rule \eqref{eqn:dw}. More precisely, we performed synaptic weight updates $w_{new} = w_{old} + \eta\,\Delta w$ for every post-pre-post spike triplet, with $t_1^{post}<t^{pre}<t_2^{post}$, where $t_1^{post}$ and $t_2^{post}$ are the spike times of two neighboring post-synaptic spikes, and $t^{pre}$ is a pre-synaptic spike time. $\eta$ is a small positive constant learning rate. The weight changes $\Delta w$ were thus given by
\begin{equation}
\Delta\,w \;=\;
        \fun{W_{3}}{\Delta t_1, \Delta t_2, w} \;=\; 
        \fun{W_{LTP}}{\Delta t_1, \Delta t_2} \;-\; 
        \left( \frac{1 - r_0}{2 \, r_0} + w \right) \fun{W_{LTD}}{\Delta t_1, \Delta t_2}
        \;+\; \frac{1}{2\,w} \;,
\label{eqn:dw-meth}
\end{equation}
with $\Delta t_1 = t_2^{post} - t^{pre}$ and $\Delta t_2 = t_2^{post} - t_1^{post}$. This is the general case for an arbitrary synaptic parameter $r_0$. Eq.~\ref{eqn:dw} shows the special case for $r_0=\frac{1}{2}$ which was used throughout this paper. $\fun{W_{LTP}}{\Delta t_1, \Delta t_2}$ and $\fun{W_{LTD}}{\Delta t_1, \Delta t_2}$ are the triplet STDP windows as depicted in Fig.~\ref{fig:model-illustration}, given by
\begin{align}
    W_{LTP}(\Delta t_1, \Delta t_2) \;=\; r_0\, \frac{\fun{\mu'}{\Delta t_1, \Delta t_2} \;+\; \frac{1}{\tau} \left(\fun{\mu}{\Delta t_1, \Delta t_2} - u_0 \right)}
    {\left(\fun{\sigma^2}{\Delta t_1, \Delta t_2}\right)' + \frac{2}{\tau}\fun{\sigma^2}{\Delta t_1, \Delta t_2}} \;,
    \label{eq:w_LTP}
\end{align}
and
\begin{equation}
    W_{LTD}(\Delta t_1, \Delta t_2) \;=\;  r_0^2\,
    \frac{ 1 }
         { {\left(\fun{\sigma^2}{\Delta t_1, \Delta t_2}\right)' + \frac{2}{\tau}\fun{\sigma^2}{\Delta t_1, \Delta t_2}} }
         \;,
    \label{eq:w_LTD}
\end{equation}
where $\fun{\mu}{\Delta t_1, \Delta t_2}$ and $\fun{\sigma^2}{\Delta t_1, \Delta t_2}$, respectively, are the estimated mean and variance of the membrane potential based on the back-propagating action potentials (see Section \ref{sec:meth-ou-bridge} for details), and $\mu'(t) = \frac{\partial}{\partial t} \mu(t)$ and $\left(\sigma^2(t)\right)' = \frac{\partial}{\partial t} \sigma^2(t)$ denote the time derivatives. In the following sections we will develop our main theoretical result to show that the synaptic weight updates \eqref{eqn:dw-meth} minimize the free energy $\frac{\partial \mathcal{F}(z,w)}{\partial w}$ of the synaptic efficacy $w$ with respect to the back-propagating action potentials $z$.

\subsection{Variational learning and free energy minimization}

The FEP proposes a specific method to approaching a state of minimum surprise. This method rests on the idea that a biological organism maintains an internal model, of its environment that allows it to reason about the external states $u$. The internal model is composed of two parts, (1) the \emph{recognition density} $\cfun{q}{u}{w}$, that describes how the external state $u$ interacts with the internal state $w$, and (2) the \emph{generative density} $\cprob{u}{z}$, that describes the dependency between external states $u$ and feedback $z$ \cite{buckley2017free}. Using this internal model the complexity of the learning problem can be approached by replacing the goal to minimize surprise directly by a variational upper bound, that allows us to split the learning problem into two parts. The theory stems from the observation that an upper bound on the surprise can be reached indirectly by employing the recognition density $q$ to \emph{guess} external states $u$, and the generative density $p$ evaluates how well the feedback $z$ agrees with the guessed external states $u$. The problem to minimize surprise is then augmented with a divergence term to also minimizing the mismatch between $q$ and $p$ through learning.

We adopted this idea and suggested to minimize an upper bound on the surprise in every synapse, given by the variational free energy $\mathcal{F}$, which is defined as
\begin{eqnarray}
\mathcal{F}(z,w) \;&=&\; -\log \prob{z} \,+\, \KL{q}{p}
\quad\geq\quad -\log \prob{z}
\label{eq:variational-bound}
\end{eqnarray}
where $\KL{q}{p}$ is the Kullback-Leibler (KL)-divergence between $q$ and $p$. The inequality follows from $\KL{q}{p} \geq 0$ for any two probability distributions $q$ and $p$, given by
\begin{align}
    \fun{\mathcal{F}}{z,w} & \;=\; -\log \prob{z} + \KL{q}{p}  \;=\; -\log \prob{z} + \expect[\cfun{q}{u}{w}]{ \log \frac{\cfun{q}{u}{w}}{\cprob{u}{z}} }\;,
    \label{eq:free-energy}
\end{align}
where $\expect[\fun{q}{u}]{ f(u) }$ denotes the expectation of some function $f(u)$ with respect to the probability density $\fun{q}{u}$. 
Learning is done in the FEP by minimizing $\fun{\mathcal{F}}{z,w}$
with respect to $w$, which can be done by gradient decent
\begin{align}
  \Delta w & \;=\; - \frac{\partial}{\partial w} \, \fun{\mathcal{F}}{z,w} \;, \label{eq:learning-rule-symbolic}
\end{align}
In the following sections we will derive the learning rule that solves this optimization problem for the case of our synapse model step by step.

\subsection{The generative density}
\label{sec:meth-ou-bridge}

In this section we formally define the generative density $\cprob{u}{z}$ which describes the dynamics of the membrane potential $u$, given the observed post-synaptic spike train $z$ back-propagating to the synapse, in Eq.~\ref{eq:free-energy}. To arrive at this result we first rewrite the dynamics of the membrane potential $u(t)$, Eq.~\eqref{eq:membrane-potential} in terms of a stochastic differential equation, by replacing the deterministic input current $y(t)$ with a stochastic one. This allows us to express the uncertainty of the synapse about $u(t)$
\begin{equation}
    d\,u \;=\; \frac{1}{\tau_m} (u_0 -  u(t)) \, dt \;+\; \sigma_0 \, d\,\mathcal{W}(t)\;,
    \label{eq:ou-process}
\end{equation}
with resting membrane potential $u_0$ and where $\sigma_0$ scales the contribution of the total stochstic input current and $\mathcal{W}(t)$ is the Wiener process. Eq.~\eqref{eq:ou-process} is an Ornstein-Uhlenbeck (OU)-process that describes the dynamics of the LIF neuron model with stochastic inputs \cite{gerstner2014neuronal}. This model is convenient because it captures the uncertainty of a synapse that is not able to observe all inputs to the post-synaptic neuron. The OU process can be solved analytically using stochastic calculus, e.g. if the process \eqref{eq:ou-process} is fixed to $u_{0}$ at time 0 it evolves according to
\begin{equation}
    u(t) \;=\; u_{0} \;+\; \sigma_{0} \int_{0}^{t} e^{-\frac{t - s}{\tau_m}} d\,\mathcal{W}(s)\,.
\end{equation}
For long observation times the OU process converges to a stationary distribution, given by a Gaussian with mean $u_0$ and variance $\sigma_0^2$, i.e. for $t \rightarrow \infty$, $u(t) \sim \cfun{\mathcal{N}}{u(t)}{u_0, \sigma_0^2}$.

The information about the spike times $z$ deflects the membrane potential from its resting state, which is expressed in the generative density $\cprob{u}{z}$. We can express the dynamics of the membrane potential given the information that the membrane potential is at the firing threshold $\vartheta$ at the firing times $t^{post}$, i.e. the constraint $u(t_1^{post})=u_{reset}$ and $u(t_2^{post})=\vartheta$ through a stochastic process with time varying mean $\mu(t)$ and variance $\sigma^2(t)$.
For our Ornstein-Uhlenbeck process model \eqref{eq:ou-process} the resulting constraint stochastic process has to fulfill the following requirements
\begin{itemize}
    \item The mean $\mu(t)$ obeys $\mu(t_1^{post})=u_{reset}$ and $\mu(t_2^{post})=\vartheta$.
    \item For $t_1^{post} < t  \leq t_2^{post}$, $\mu(t)$ approaches the resting potential $u_0$ asymptotically.
    \item The variance $\sigma^2(t)$ approaches 0 when close to the firing times $t_1^{post}$ and $t_2^{post}$.
    \item For $t_1^{post} < t  \leq t_2^{post}$, $\sigma^2(t)$ approaches the variance $\sigma^2_0$ of the stationary distribution asymptotically.
    \item The functions $\mu(t)$ and $\sigma^2(t)$ are smooth and follow the LIF dynamics with time constant $\tau_m$.
\end{itemize}
In Appendix~\ref{sec:apx-generative-density} we determine a stochastic dynamics that fulfills the above constraints. Using this result, for any neighboring postsynaptic spike pair $(t_1^{post}, t_2^{post})$ and time point $t$ with $t_1^{post} < t  \leq t_2^{post}$ we describe the dynamics of $u(t)$ using its mean $\mu(t)$ and variance function $\sigma^2(t)$, given by
\begin{align}
    \mu(t) \;=\; \expect[]{u(t)} & \;=\; u_0
    \;+\; (u_{reset} - u_0)
        \frac{e^{\frac{\Delta t_1}{\tau_m}} -
              e^{-\frac{\Delta t_1}{\tau_m}}}
             {e^{\frac{\Delta t_2}{\tau_m}} -
              e^{-\frac{\Delta t_2}{\tau_m}} }
    \;+\; (\vartheta - u_0)
        \frac{e^{\frac{\Delta t_2 - \Delta t_1}{\tau_m}} - e^{\frac{ \Delta t_1 - \Delta t_2}{\tau_m}}}
             {e^{\frac{\Delta t_2}{\tau_m}} -
              e^{-\frac{\Delta t_2}{\tau_m}} }
    \label{eq:mu_ou}
\end{align}
and
\begin{equation}
    \sigma^2(t) \;=\; \expect[]{u^2(t)} - \expect[]{u(t)} \;=\; \sigma_{0}^2\; \frac{1}
         {1 \,+\, \gamma \left( 
         e^{\frac{\Delta t_1 - \Delta t_2}{\tau_m}} +
         e^{-\frac{\Delta t_1}{\tau_m}} \right) } \;,
    \label{eq:sigma_ou}
\end{equation}
where $\Delta t_1 = t_2^{post} - t$, $\Delta t_2 = t_2^{post} - t_1^{post}$ and $\gamma$ is a constant that scales the slope of the variance function. In other words, the dynamics of the membrane potential subject to the constraint $u(t_1^{post})=u_{reset}$ and $u(t_2^{post})=\vartheta$ are described by a stochastic process with mean $\mu(t)$ and variance $\sigma^2(t)$. The membrane potential mean and variance functions \eqref{eq:mu_ou} and \eqref{eq:sigma_ou} are piece-wise defined for all postsynaptic spike intervals $(t_{n}^{post}, t_{n+1}^{post})$. The membrane dynamics during each interval are statistically independent of each other due to the resetting behavior of the neuron model.

Using this result, we define the generative density for every time point $t$, as $\cprob{u(t)}{z} = \cfun{\mathcal{N}}{u(t)}{\mu(t), \sigma^2(t)}$. Since the mean \eqref{eq:mu_ou} and variance \eqref{eq:sigma_ou} functions only depend on the relative spike timing $\Delta t_1$ and $\Delta t_2$ the learning rule Eq.~\ref{eqn:dw-meth} can be expressed through triplet STDP kernels, where $\fun{\mu}{\Delta t_1, \Delta t_2}$ and $\fun{\sigma^2}{\Delta t_1, \Delta t_2}$, respectively, denote $\mu$ and $\sigma^2$ evaluated at the pre-synaptic spike time $t^{pre}$.

\subsection{Derivation of the learning rule}
\label{sec:meth-learning-rule}

Finally we make use of the result from Section~\ref{sec:apx-recognition-density} to rewrite the free energy. We exploit here that that the OU process model \eqref{eq:ou-process} suggests a one-to-one relation between synaptic inputs $y(t)$ and somatic membrane potentials $u(t)$, that is, for a given $y(t)$ we can determine $u(t)$ through a deterministic function. Using this Eq.~\eqref{eq:learning-rule-symbolic} becomes
\begin{align}
  \Delta w & \;=\;
  - \frac{\partial}{\partial w} \fun{\mathcal{F}}{z,w} 
  \;=\; \frac{\partial}{\partial w} \expect[\cfun{q}{u(t)}{w}]{ \log \frac{\cprob{u(t)}{z}}{\cfun{q}{u(t)}{w}} }
  \;=\; \frac{\partial}{\partial w} \expect[\cfun{q}{y(t)}{w}]{ \log \frac{\cprob{y(t)}{z}}{\cfun{q}{y(t)}{w}} }\;.
  \label{eq:learning-rule-expanded}
\end{align}
This last result is useful, because the generative model established in Section~\ref{sec:meth-ou-bridge} allows us to express the posterior distribution $\cfun{q}{y(t)}{w}$ in closed form. In Appendix~\ref{sec:apx-recognition-density} we show in detail that a synaptic current $y(t) \sim \cfun{\mathcal{N}}{y(t)}{a(t),b(t)}$ enables a synapse to realize a somatic membrane potential $u(t)$ that obeys the stochastic processes with mean $\mu(t)$ and variance  $\sigma^2(t)$, if
\begin{equation}
\begin{split}
a(t) &\;=\; \mu'(t) + \frac{1}{\tau} \left(\mu(t) - u_0 \right)\;, \\[3mm]
b(t) &\;=\; \left(\sigma^2(t)\right)' + \frac{2}{\tau}\sigma^2(t)\;,
\end{split}
\label{eqn:a-b}
\end{equation}
such that $\cprob{y(t)}{z} = \cfun{\mathcal{N}}{y(t)}{a(t),b(t)}$, where $\mu(t)$ and $\sigma^2(t)$ are as defined for the constraint stochastic process as defined above. This result is also the basis for the simulations presented in Fig.~\ref{fig:bridge-model}C,D. Furthermore, the stochastic synapse model Eq.~\eqref{eq:synaptic-current} suggests that at the time points of pre-synaptic firing $t^{pre}$ the amplitudes of synaptic currents follow a Gaussian distribution $\cfun{q}{y}{w} = \cfun{\mathcal{N}}{y}{r_0\,w, s_0\,w}$.

\begin{figure}
	\centering
	\includegraphics[width=\textwidth]{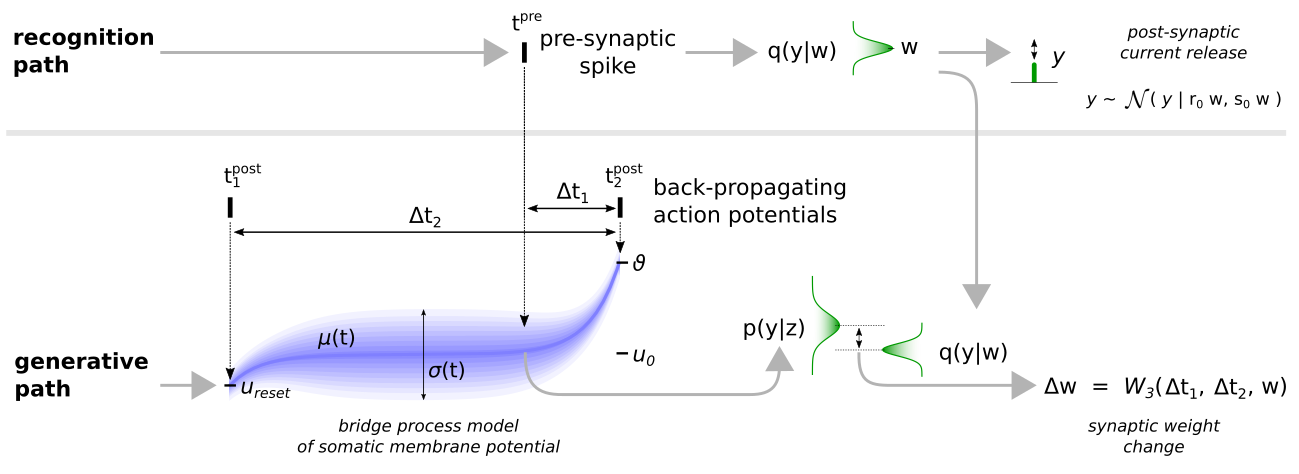}
	\caption{
	\textbf{Illustration of the rationale behind the learning rule Eq.~\eqref{eqn:dw-meth}.}
		The weight updates can be split into two paths. The recognition path uses the recognition density $\cfun{q}{y}{w}$ to generate a synaptic current $y$. In the generative path this result is compared to the posterior distribution according to the stochastic bridge model of the membrane potential.
		\label{fig:fep-rule-illustration}
    }
\end{figure}

To construct the term $\frac{\partial}{\partial w} \expect[\cfun{q}{y(t)}{w}]{ \log \frac{\cprob{y(t)}{z}}{\cfun{q}{y(t)}{w}} }$ of Eq.~\eqref{eq:learning-rule-expanded} we use the result from Section~\ref{sec:meth-ou-bridge} to get
\begin{align}
\frac{\partial}{\partial w}
   & \expect[\cfun{q}{y(t)}{w}]{ \log \frac{\cprob{y(t)}{z}}{\cfun{q}{y(t)}{w}} } \nonumber \\
   & \;=\;
\frac{\partial}{\partial w}
   \expect[\cfun{q}{y(t)}{w}]{ - \frac{1}{2} \log \left( 2\pi\, b(t) \right) \;-\; \frac{\left( y(t) - a(t) \right)^2}{2 b(t)} } \;+\; \frac{1}{2} \frac{\partial}{\partial w} \log\left( 2\,\pi\,e\,\sigma_w^2(t)\right) \nonumber \\
   & \;=\;
\frac{\partial}{\partial w}
   \expect[\cfun{q}{y(t)}{w}]{ \frac{2\,y(t)\,a(t) \;-\; y^2(t) }{ 2\,b(t)} } \;+\; \frac{1}{2} \, \frac{1}{\sigma_w^2(t)} \, \frac{\partial}{\partial w} \sigma_w^2(t) \nonumber \\
   & \;=\;
    \left( \frac{a(t)}{b(t)} \right)
    \frac{\partial}{\partial w} \expect[\cfun{q}{y(t)}{w}]{ y(t) } \;-\;
    \frac{1}{2} \,
    \left( \frac{1}{b(t)} \right)
    \frac{\partial}{\partial w} \expect[\cfun{q}{y(t)}{w}]{ y^2(t) } \;+\; \frac{1}{2} \, \frac{1}{\sigma_w^2(t)} \, \frac{\partial}{\partial w} \sigma_w^2(t) \;.
    \label{eq:ltp-ltd-derivation}
\end{align}
By plugging in Eq.~\eqref{eq:mu_ou} and \eqref{eq:sigma_ou} we recover the LTP and LTD term in Eq.~\eqref{eqn:dw-meth}.
The rational underlying equation \eqref{eq:ltp-ltd-derivation} is illustrated in Fig.~\ref{fig:fep-rule-illustration}.

Using $\expect[\cfun{q}{y(t)}{w}]{ u(t) } = \mu_w(t)$ and $\expect[\cfun{q}{y(t)}{w}]{ u^2(t) } = \mu_w^2(t) + \sigma_w^2(t)$, we get
\begin{align}
\frac{\partial}{\partial w}
   & \expect[\cfun{q}{y(t)}{w}]{ \log \frac{\cprob{y(t)}{z}}{\cfun{q}{y(t)}{w}} } \;=\;
    \left( \frac{a(t) - \mu_w(t)}{b(t)} \right)
    \frac{\partial}{\partial w} \mu_w(t) \;-\;
    \frac{1}{2} \,
    \left( \frac{1}{b(t)} - \frac{1}{\sigma_w^2(t)} \right)
    \frac{\partial}{\partial w} \sigma_w^2(t) \;.
    \label{eq:ltp-ltd-general}
\end{align}
Finally, using Eq.~\ref{eq:synaptic-current} we identify $\mu_w(t)$ and $\sigma^2_w(t)$ get the final result for any $t$ at the pre-synaptic firing times $t^{pre}$
\begin{align}
\frac{\partial}{\partial w}
   \expect[\cfun{q}{y(t)}{w}]{ \log \frac{\cprob{y(t)}{z}}{\cfun{q}{y(t)}{w}} } &\;=\;
    \left( \frac{a(t) - r_0\,w}{b(t)} \right)
    r_0 \;-\;
    \frac{1}{2} \,
    \left( \frac{1}{b(t)} - \frac{1}{w\,r_0\,(1-r_0)} \right)
    \,r_0\,(1-r_0) \\
    &\;=\;
    r_0\,\frac{a(t)}{b(t)}
    \;-\; r_0\,\frac{1}{b(t)}
    \left( \frac{1 - r_0}{2} + r_0\,w \right)  \;+\;
    \frac{1}{2\,w} \;,
    \label{eq:ltp-ltd-final}
\end{align}
which is identical to the result in Eq.~\eqref{eqn:dw-meth}.

\subsection{Numerical simulations}

All simulations were performed in Python (3.8.5) using the Euler method to approximate the solution of the stochastic differential equations with a fixed time step of 1\,ms. Post-Synaptic currents were created as described in Eq.~\eqref{eq:synaptic-current} where Dirac delta pulses were approximated by 1\,ms rectangular pulses and truncated at zero to avoid negative currents. If not stated otherwise we used the following parameters. The synaptic current release probability $r_0$ was $\frac{1}{2}$. In Eq.~\eqref{eq:membrane-potential} the membrane time constant $\tau_m$ was 30\,ms, the resting potential $u_{0}$ was -70\,mV and the membrane resistance $R$ was 10\,M$\Omega$. The firing threshold $\vartheta$ was -55\,mV and  $u_{reset}$ was -75\,mV. For variance function Eq.~\eqref{eq:sigma_ou} we used $\gamma=10$. The learning rate $\eta$ was $10^{-5}$.

\subsubsection{Details to Figure~\ref{fig:pattern-classification}}

Rate patterns in Fig.~\ref{fig:pattern-classification} were generated by randomly drawing values from a beta distribution ($\alpha=0.1$, $\beta=0.8$) for each input channel and multiplying these values with the maximum rate of 50\,~Hz. From these rate patterns individual Poisson spike trains were drawn for every pattern presentation. During the learning phase the output neurons were clamped to fire at 50~Hz during presentation of the preferred stimulus and remain silent otherwise.

In Fig.~\ref{fig:pattern-classification}C we used a simple threshold adaptation mechanism to control the output rate of the neurons for the unsupervised case. Individual firing thresholds $\vartheta$ where used for every neuron. Thresholds were decreased by a value of $10^{-5}$\,mV in every millisecond and increased by $10^{-3}$\,mV after every output spike. 
In Fig.~\ref{fig:pattern-classification}E we set the synaptic reliability parameter $r_0$ in Eq.~\eqref{eq:synaptic-current} to 1 to disable synaptic noise and trained a new network to create the \emph{without noise} results.

\subsubsection{Details to Figure~\ref{fig:closed-loop}}

In Fig.~\ref{fig:closed-loop} we used a feed-forward network with 200 feedback neurons and 100 action neurons. Preferred positions of the feedback neurons where scattered uniformly over the unit cube and firing rates were set scaled by the euclidean distance between agent position and preferred position. Action neurons were randomly assigned to preferred directions. Agent position were updated after every 50\,ms time frame by adding the decoded position offset provided by the action neurons.

\clearpage

\appendix

\renewcommand{\theequation}{A.\arabic{equation}}
\counterwithin*{equation}{section}

\section{Appendix}

\subsection{Derivation of the recognition density}
\label{sec:apx-recognition-density}

Here we derive the recognition density $\cfun{q}{u}{w}$ for our synapse model. We will show that $\cfun{q}{u}{w}$ can be expressed as a Gaussian distribution with time-varying mean and variance functions $\mu_w(t)$ and $\sigma_w^2(t)$.
This result can be obtained by stochastic integration, but to keep this paper self-contained we provide a simple proof here. We start by considering a general drift-diffusion process and then show the special case of the LIF neuron dynamics step by step.

For the general case we consider the Fokker-Planck equation that governs the probability density function $p(x,t)$ of a stochastic process $x$ at time $t$, with drift $A(x,t)$ and diffusion $B(x,t)$
\begin{equation}
    \frac{\partial}{\partial t} p(x,t) \;=\; -\,\frac{\partial}{\partial x} \left( A(x,t) \cdot p(x,t) \right) \;+\; \frac{1}{2} \, \frac{\partial^2}{\partial x^2} \left( B(x,t) \cdot p(x,t) \right)\;.
    \label{eq:fpe-general}
\end{equation}
We treat here the case that $p(x,t)$ is a Gaussian distribution with time-varying mean $\mu(t)$ and variance $\sigma^2(t)$ functions, i.e.$p(x,t) = \cfun{\mathcal{N}}{x}{\mu(t),\sigma^2(t)}$ at any time point $t$, to get
\begin{equation*}
    \frac{\partial}{\partial t} p(x,t) \;=\; p(x,t)\,\left( \mu'(t) \frac{x - \mu(t)}{\sigma^2(t)} \;+\; \frac{1}{2} \left(\sigma^2(t)\right)' \left( \frac{(x-\mu(t))^2}{\sigma^4(t)} \,-\, \frac{1}{\sigma^2(t)} \right) \right) \;,
\end{equation*}
where $\mu'(t) = \frac{\partial}{\partial t} \mu(t)$ and $\left(\sigma^2(t)\right)' = \frac{\partial}{\partial t} \sigma^2(t)$. Furthermore
\begin{align*}
    &\frac{\partial}{\partial x} \left( A(x,t) \cdot p(x,t) \right) \;=\; p(x,t) \left(  \frac{\partial}{\partial x} A(x,t) \;-\; A(x,t) \frac{x-\mu(t)}{\sigma^2(t)} \right)\\[3mm]
    &\text{and} \\[3mm]
    &\frac{\partial^2}{\partial x^2} \left( B(x,t) \cdot p(x,t) \right) \;=\; p(x,t) \left(  \frac{\partial^2}{\partial x^2} B(x,t) \;-\; 2 \frac{\partial}{\partial x} B(x,t) \frac{x-\mu(t)}{\sigma^2(t)} \;+\; B(x,t) \left( \frac{(x-\mu(t))^2}{\sigma^4(t)} \,-\, \frac{1}{\sigma^2(t)} \right)  \right)\;.
\end{align*}
Therefore, by plugging these results back into the Fokker-Planck equation \eqref{eq:fpe-general}, we find the condition that has to be satisfied for functions $A(x,t)$ and $B(x,t)$ to be given by
\begin{eqnarray}
    && \mu'(t) \frac{x - \mu(t)}{\sigma^2(t)} \;+\; \frac{1}{2} \left(\sigma^2(t)\right)' \left( \frac{(x-\mu(t))^2}{\sigma^4(t)} \,-\, \frac{1}{\sigma^2(t)} \right) \quad \overset{!}{=} \label{eq:fpe-cond} \\
    && \quad  A(x,t) \frac{x-\mu(t)}{\sigma^2(t)} \;-\; \frac{\partial}{\partial x} A(x,t) \;+\; \frac{1}{2} \frac{\partial^2}{\partial x^2} B(x,t) \;-\; \frac{\partial}{\partial x} B(x,t) \frac{x-\mu(t)}{\sigma^2(t)} \;+\; \frac{1}{2} B(x,t) \left( \frac{(x-\mu(t))^2}{\sigma^4(t)} \,-\, \frac{1}{\sigma^2(t)} \right)\;. \nonumber
\end{eqnarray}
Clearly, the choice $A(x,t) = \mu'(t)$, $B(x,t) = \left(\sigma^2(t)\right)'$ satisfies this condition for any differentiable functions $\mu(t)$ and $\sigma^2(t)$.

To arrive at the final result we replace the general drift-diffusion dynamics with the special case of a leaky integrator with finite integration time constant $\tau$ using the ansatz $A(x,t) = \frac{1}{\tau} \left(x_0 - x\right) + a(t)$ and $B(x,t) = b(t)$. In this case, we can make condition \eqref{eq:fpe-cond} satisfied if $\mu'(t) = \frac{1}{\tau} \left(x_0 - \mu(t)\right) + a(t)$ and $\left(\sigma^2(t)\right)' = b(t) - \frac{2}{\tau} \sigma^2(t)$. This can be verified by plugging this result back into Eq.~\eqref{eq:fpe-cond}
\begin{eqnarray*}
    && \left( \frac{1}{\tau} \left(x_0 - \mu(t)\right) + a(t) \right) \frac{x - \mu(t)}{\sigma^2(t)} \;+\; \frac{1}{2} \left( b(t) - \frac{2}{\tau} \sigma^2(t) \right) \left( \frac{(x-\mu(t))^2}{\sigma^4(t)} \,-\, \frac{1}{\sigma^2(t)} \right) \quad \overset{!}{=} \\
    && \qquad  \left( \frac{1}{\tau} \left(x_0 - x\right) + a(t) \right) \frac{x-\mu(t)}{\sigma^2(t)} \;+\; \frac{1}{\tau} \;+\; \frac{1}{2} b(t) \left( \frac{(x-\mu(t))^2}{\sigma^4(t)} \,-\, \frac{1}{\sigma^2(t)} \right)\;,
\end{eqnarray*}
from which the equality follows
\begin{eqnarray*}
    \leftrightarrow && \frac{1}{\tau} \left(x_0 - \mu(t)\right) \frac{x - \mu(t)}{\sigma^2(t)} \;-\; \frac{1}{\tau} \left( \frac{(x-\mu(t))^2}{\sigma^2(t)} \,-\, 1 \right) \quad \overset{!}{=} \quad  \frac{1}{\tau} \left(x_0 - x\right) \frac{x-\mu(t)}{\sigma^2(t)} \;+\; \frac{1}{\tau} \; \\
    \leftrightarrow &&  \left(x_0 - \mu(t)\right) \frac{x - \mu(t)}{\sigma^2(t)} \;-\; \frac{(x-\mu(t))^2}{\sigma^2(t)} \quad \overset{!}{=} \quad \left(x_0 - x\right) \frac{x-\mu(t)}{\sigma^2(t)} \; \\
    \leftrightarrow &&  \left(x_0 - x\right) \frac{x-\mu(t)}{\sigma^2(t)} \quad \overset{!}{=} \quad \left(x_0 - x\right) \frac{x-\mu(t)}{\sigma^2(t)} \qquad \checkmark \; \\
\end{eqnarray*}
This proofs that a stochastic process $x$ with $p(x,t) = \cfun{\mathcal{N}}{x}{\mu(t),\sigma^2(t)}$, $\mu'(t) = \frac{1}{\tau} \left(x_0 - \mu(t)\right) + a(t)$ and $\left(\sigma^2(t)\right)' = b(t) - \frac{2}{\tau} \sigma^2(t)$ is realized by a drift $A(x,t) = \frac{1}{\tau} \left(x_0 - x\right) + a(t)$ and diffusion $B(x,t) = b(t)$.
Equivalently, any process $x$ with mean $\mu(t)$ and variance $\sigma^2(t)$ can be realized if
\begin{equation}
\begin{split}
a(t) &\;=\; \mu'(t) + \frac{1}{\tau} \left(\mu(t) - x_0 \right)\;, \\[3mm]
b(t) &\;=\; \left(\sigma^2(t)\right)' + \frac{2}{\tau}\sigma^2(t)\;,
\end{split}
\label{eqn:apx-a-b}
\end{equation}
and $b(t) \geq 0$ can be satisfied for all $t$. This last result was used in the main text to derive the learning rule \eqref{eqn:dw-meth}.

Furthermore, by integration of this last result we find that any integrable functions $a(t)$ and $b(t) > 0$ yield the following dynamics for the stochastic process $x$
\begin{equation}
\begin{split}
    \mu(t) \;&=\; x_0 \;+\; e^{-\frac{t}{\tau}} \int_0^t  e^{\frac{s}{\tau}} a(s) \; d\,s \\[3mm]
    \sigma^2(t) \;&=\; e^{-\frac{2t}{\tau}} \int_0^t  e^{\frac{2s}{\tau}} b(s) \; d\,s\;.
\end{split}
\end{equation}
For $a(t) = 0$ and $b(t) = b$ (constant) we recover the Ornstein-Uhlenbeck process dynamics. 

For our synapse model drift and diffusion are governed by the stochastic current release, which can be reflected in the above model by using $a(t) = r_0\,R\,w \sum_f \delta(t - t^{(f)})$ and $b(t) = r_0\,(1-r_0)\,R\,w \sum_f \delta(t - t^{(f)})$, where $\delta(t)$ is the Dirac delta function. Using this we identify the recognition density $q_w(u(t)) = \cfun{\mathcal{N}}{u(t)}{\mu_w(t), \sigma_w^2(t)}$, with
\begin{equation}
\begin{split}
    \mu_w(t) \;&=\; u_0 \;+\; r_0\,R\,w\,\sum_f\fun{\epsilon}{t-t^{(f)}} \\[3mm]
    \sigma_w^2(t) \;&=\; r_0\,(1-r_0)\,R^2\,w\,\sum_f\fun{\epsilon}{2\,t-2\,t^{(f)}} \;,
    \label{eq:recognition-density-mean-var}
\end{split}
\end{equation}
where $\fun{\epsilon}{t} = e^{-\frac{t}{\tau_m}} \Theta(t)$ and $\Theta(t)$ is the Heaviside step function.

\subsection{Details to the generative density}
\label{sec:apx-generative-density}

The generative density $\cprob{u}{z}$ describes the evolution of the external world and the uncertainty of the synapse based on the provided feedback information $z$. In our model the feedback is given by the back-propagating action potentials which precisely determine the $u$ at the spike times. At all other times the precise value of the membrane potential is unknown and this uncertainty should be reflected in the generative density.

We use a Gaussian process model of the external world, such that the generative density is given by
\begin{equation}
    \cprob{u(t)}{z} \;=\; \cfun{\mathcal{N}}{u(t)}{\mu(t), \sigma^2(t)}\;.
\end{equation}

In principle we can assume any function $\mu(t)$ and $\sigma^2(t)$ and the learning rule Eq.~\ref{eqn:dw} will strive to best approximate its dynamics. However, a reasonable choice will obey the constraints imposed by the neuron and synapse dynamics, e.g. the membrane time constant and the firing mechanism and resetting behavior of the neuron.

The LIF neuron implies OU process dynamics of the membrane potential. Given the information that the membrane potential is at the firing threshold $\vartheta$ at the firing times $t^{post}$, i.e. the constraint $u(t_1^{post})=u_{reset}$ and $u(t_2^{post})=\vartheta$, the OU process can be solved explicitly. The solution to this double constraint stochastic process is the OU-bridge process \cite{corlay2013properties,szavits2015inequivalence}. For any neighboring postsynaptic spike pair $(t_1^{post}, t_2^{post})$ and time point $t$ with $t_1^{post} < t  \leq t_2^{post}$ we can describe the dynamics of $u(t)$ using its mean $\mu(t)$, given by
\begin{align}
    \mu(t) \;=\; \expect[]{u(t)} & \;=\; u_0
    \;+\; (u_{reset} - u_0)
        \frac{\sinh\left( \frac{t_2^{post}-t}{\tau_m} \right)}
             {\sinh\left( \frac{t_2^{post} - t_1^{post}}{\tau_m} \right)}
    \;+\; (\vartheta - u_0)
        \frac{\sinh\left( \frac{t - t_1^{post}}{\tau_m} \right)}
             {\sinh\left( \frac{t_2^{post} - t_1^{post}}{\tau_m} \right)}\;,
    \label{eq:mu_ou_apx}
\end{align}
where $\sinh(t) = \frac{e^t - e^{-t}}{2}$ is the hyperbolic sine function. This function describes the asymptotic approach to the resting potential $u_0$ and the slope towards the firing threshold $\vartheta$. We used this function to determine the generative density.

The OU-bridge process model also provides a solution for the variance function $\sigma^2(t)$, given by
\begin{equation}
    \sigma^2(t) \;=\; \expect[]{u^2(t)} - \expect[]{u(t)} \;=\; \sigma_{0}^2\; \frac{\sinh\left(\frac{t_2^{post}-t}{\tau_m}\right) \,\times\,
          \sinh\left(\frac{t-t_1^{post}}{\tau_m}\right)}
         {\sinh\left(\frac{t_2^{post}-t_1^{post}}{\tau_m}\right)}\;.
    \label{eq:sigma_ou_bridge}
\end{equation}
This variance function has a very steep slope to approach the firing threshold. It is so steep in fact that it would require us to have negative variance for the synaptic current $y$ close to the firing times to realize the process Eq.~\eqref{eqn:apx-a-b}. This has no meaningful physical interpretation and therefore we used the slower process that better reflects the asymptotic behavior of a leaky integrator, given by
\begin{equation}
    \sigma^2(t) \;=\; \sigma_{0}^2\; \frac{\gamma}
         {\gamma + 
         e^{\frac{t_1^{post} - t}{\tau_m}} +
         e^{\frac{t - t_2^{post}}{\tau_m}}} \;.
    \label{eq:sigma_ou_apx}
\end{equation}


\begin{thebibliography}{}

\bibitem[Aitchison et~al., 2021]{aitchison2021synaptic}
Aitchison, L., Jegminat, J., Menendez, J.~A., Pfister, J.-P., Pouget, A., and
  Latham, P.~E. (2021).
\newblock Synaptic plasticity as bayesian inference.
\newblock {\em Nature Neuroscience}, pages 1--7.

\bibitem[Aitchison et~al., 2014]{aitchison2014probabilistic}
Aitchison, L., Pouget, A., and Latham, P.~E. (2014).
\newblock Probabilistic synapses.
\newblock {\em arXiv preprint arXiv:1410.1029}.

\bibitem[Barascud et~al., 2016]{barascud2016brain}
Barascud, N., Pearce, M.~T., Griffiths, T.~D., Friston, K.~J., and Chait, M.
  (2016).
\newblock Brain responses in humans reveal ideal observer-like sensitivity to
  complex acoustic patterns.
\newblock {\em Proceedings of the National Academy of Sciences},
  113(5):E616--E625.

\bibitem[Bastos et~al., 2012]{BastosETAL:12}
Bastos, A.~M., Usrey, W.~M., Adams, R.~A., Mangun, G.~R., Fries, P., and
  Friston, K.~J. (2012).
\newblock Canonical microcircuits for predictive coding.
\newblock {\em Neuron}, 76(4):695--711.

\bibitem[Borst, 2010]{borst2010low}
Borst, J. G.~G. (2010).
\newblock The low synaptic release probability in vivo.
\newblock {\em Trends in neurosciences}, 33(6):259--266.

\bibitem[Brea et~al., 2013]{BreaETAL:13}
Brea, J., Senn, W., and Pfister, J.-P. (2013).
\newblock Matching recall and storage in sequence learning with spiking neural
  networks.
\newblock {\em The Journal of Neuroscience}, 33(23):9565--9575.

\bibitem[Buckley et~al., 2017]{buckley2017free}
Buckley, C.~L., Kim, C.~S., McGregor, S., and Seth, A.~K. (2017).
\newblock The free energy principle for action and perception: A mathematical
  review.
\newblock {\em Journal of Mathematical Psychology}, 81:55--79.

\bibitem[Buesing and Maass, 2008]{buesing2008simplified}
Buesing, L. and Maass, W. (2008).
\newblock Simplified rules and theoretical analysis for information bottleneck
  optimization and pca with spiking neurons.
\newblock In {\em Advances in Neural Information Processing Systems}, pages
  193--200.

\bibitem[Buesing and Maass, 2010]{buesing2010spiking}
Buesing, L. and Maass, W. (2010).
\newblock A spiking neuron as information bottleneck.
\newblock {\em Neural computation}, 22(8):1961--1992.

\bibitem[Caporale and Dan, 2008]{caporale2008spike}
Caporale, N. and Dan, Y. (2008).
\newblock Spike timing--dependent plasticity: a hebbian learning rule.
\newblock {\em Annu. Rev. Neurosci.}, 31:25--46.

\bibitem[Chalk et~al., 2018]{chalk2018toward}
Chalk, M., Marre, O., and Tka{\v{c}}ik, G. (2018).
\newblock Toward a unified theory of efficient, predictive, and sparse coding.
\newblock {\em Proceedings of the National Academy of Sciences},
  115(1):186--191.

\bibitem[Clopath et~al., 2010]{clopath2010connectivity}
Clopath, C., B{\"u}sing, L., Vasilaki, E., and Gerstner, W. (2010).
\newblock Connectivity reflects coding: a model of voltage-based stdp with
  homeostasis.
\newblock {\em Nature neuroscience}, 13(3):344.

\bibitem[Clopath and Gerstner, 2010]{clopath2010voltage}
Clopath, C. and Gerstner, W. (2010).
\newblock Voltage and spike timing interact in stdp -- a unified model.
\newblock {\em Frontiers in synaptic neuroscience}, 2:25.

\bibitem[Corlay, 2013]{corlay2013properties}
Corlay, S. (2013).
\newblock Properties of the ornstein-uhlenbeck bridge.
\newblock {\em arXiv preprint arXiv:1310.5617}.

\bibitem[Dan and Poo, 2004]{dan2004spike}
Dan, Y. and Poo, M.-m. (2004).
\newblock Spike timing-dependent plasticity of neural circuits.
\newblock {\em Neuron}, 44(1):23--30.

\bibitem[Davies et~al., 2018]{davies2018loihi}
Davies, M., Srinivasa, N., Lin, T.-H., Chinya, G., Cao, Y., Choday, S.~H.,
  Dimou, G., Joshi, P., Imam, N., Jain, S., et~al. (2018).
\newblock Loihi: A neuromorphic manycore processor with on-chip learning.
\newblock {\em Ieee Micro}, 38(1):82--99.

\bibitem[Deneve, 2008]{deneve2008bayesian}
Deneve, S. (2008).
\newblock Bayesian spiking neurons ii: learning.
\newblock {\em Neural computation}, 20(1):118--145.

\bibitem[DJ~Rezende, 2014]{RezendeETAL:14}
DJ~Rezende, S~Mohamed, D.~W. (2014).
\newblock Stochastic backpropagation and approximate inference in deep
  generative models.
\newblock {\em International Conference on Machine Learning}.

\bibitem[DJ~Rezende, 2011]{RezendeETAL:11}
DJ~Rezende, D~Wierstra, W.~G. (2011).
\newblock Variational learning for recurrent spiking networks.
\newblock {\em Advances in Neural Information Processing Systems}, pages
  136--144.

\bibitem[Driscoll et~al., 2017]{driscoll2017dynamic}
Driscoll, L.~N., Pettit, N.~L., Minderer, M., Chettih, S.~N., and Harvey, C.~D.
  (2017).
\newblock Dynamic reorganization of neuronal activity patterns in parietal
  cortex.
\newblock {\em Cell}, 170(5):986--999.

\bibitem[Faisal et~al., 2008]{faisal2008noise}
Faisal, A.~A., Selen, L.~P., and Wolpert, D.~M. (2008).
\newblock Noise in the nervous system.
\newblock {\em Nature reviews neuroscience}, 9(4):292--303.

\bibitem[Feinstein, 1986]{feinstein1986relating}
Feinstein, D.~I. (1986).
\newblock {\em Relating thermodynamics to information theory: the equality of
  free energy and mutual information}.
\newblock PhD thesis, California Institute of Technology.

\bibitem[Friston, 2005]{friston2005theory}
Friston, K. (2005).
\newblock A theory of cortical responses.
\newblock {\em Philosophical transactions of the Royal Society B: Biological
  sciences}, 360(1456):815--836.

\bibitem[Friston, 2008]{Friston:08}
Friston, K. (2008).
\newblock Variational filtering.
\newblock {\em NeuroImage}, 41(3):747--766.

\bibitem[Friston, 2010]{friston2010free}
Friston, K. (2010).
\newblock The free-energy principle: a unified brain theory?
\newblock {\em Nature reviews neuroscience}, 11(2):127.

\bibitem[Friston and Ao, 2012]{friston2012attractors}
Friston, K. and Ao, P. (2012).
\newblock Free energy, value, and attractors.
\newblock {\em Computational and mathematical methods in medicine}, 2012.

\bibitem[Friston et~al., 2014]{friston2014anatomy}
Friston, K., Schwartenbeck, P., FitzGerald, T., Moutoussis, M., Behrens, T.,
  and Dolan, R.~J. (2014).
\newblock The anatomy of choice: dopamine and decision-making.
\newblock {\em Philosophical Transactions of the Royal Society B: Biological
  Sciences}, 369(1655):20130481.

\bibitem[Gerstner et~al., 2014]{gerstner2014neuronal}
Gerstner, W., Kistler, W.~M., Naud, R., and Paninski, L. (2014).
\newblock {\em Neuronal dynamics: From single neurons to networks and models of
  cognition}.
\newblock Cambridge University Press.

\bibitem[Gjorgjieva et~al., 2011]{gjorgjieva2011triplet}
Gjorgjieva, J., Clopath, C., Audet, J., and Pfister, J.-P. (2011).
\newblock A triplet spike-timing--dependent plasticity model generalizes the
  bienenstock--cooper--munro rule to higher-order spatiotemporal correlations.
\newblock {\em Proceedings of the National Academy of Sciences},
  108(48):19383--19388.

\bibitem[Gontier and Pfister, 2020]{gontier2020identifiability}
Gontier, C. and Pfister, J.-P. (2020).
\newblock Identifiability of a binomial synapse.
\newblock {\em Frontiers in computational neuroscience}, 14:86.

\bibitem[Graupner and Brunel, 2012]{graupner2012calcium}
Graupner, M. and Brunel, N. (2012).
\newblock Calcium-based plasticity model explains sensitivity of synaptic
  changes to spike pattern, rate, and dendritic location.
\newblock {\em Proceedings of the National Academy of Sciences},
  109(10):3991--3996.

\bibitem[Isomura and Friston, 2018]{isomura2018vitro}
Isomura, T. and Friston, K. (2018).
\newblock In vitro neural networks minimise variational free energy.
\newblock {\em Scientific reports}, 8(1):16926.

\bibitem[Isomura et~al., 2015]{isomura2015cultured}
Isomura, T., Kotani, K., and Jimbo, Y. (2015).
\newblock Cultured cortical neurons can perform blind source separation
  according to the free-energy principle.
\newblock {\em PLoS computational biology}, 11(12):e1004643.

\bibitem[Isomura et~al., 2016]{isomura2016linking}
Isomura, T., Sakai, K., Kotani, K., and Jimbo, Y. (2016).
\newblock Linking neuromodulated spike-timing dependent plasticity with the
  free-energy principle.
\newblock {\em Neural computation}, 28(9):1859--1888.

\bibitem[Jensen et~al., 2019]{jensen2019multiplex}
Jensen, T.~P., Zheng, K., Cole, N., Marvin, J.~S., Looger, L.~L., and Rusakov,
  D.~A. (2019).
\newblock Multiplex imaging relates quantal glutamate release to presynaptic ca
  2+ homeostasis at multiple synapses in situ.
\newblock {\em Nature communications}, 10(1):1--14.

\bibitem[Kanai et~al., 2015]{kanai2015cerebral}
Kanai, R., Komura, Y., Shipp, S., and Friston, K. (2015).
\newblock Cerebral hierarchies: predictive processing, precision and the
  pulvinar.
\newblock {\em Philosophical Transactions of the Royal Society B: Biological
  Sciences}, 370(1668):20140169.

\bibitem[Kappel et~al., 2014]{KappelETAL:14}
Kappel, D., Nessler, B., and Maass, W. (2014).
\newblock Stdp installs in winner-take-all circuits an online approximation to
  hidden markov model learning.
\newblock {\em PLoS Computational Biology}, 10(3):e1003511.

\bibitem[Katz, 1971]{katz1971quantal}
Katz, B. (1971).
\newblock Quantal mechanism of neural transmitter release.
\newblock {\em Science}, 173(3992):123--126.

\bibitem[Kiebel and Friston, 2011]{kiebel2011free}
Kiebel, S.~J. and Friston, K.~J. (2011).
\newblock Free energy and dendritic self-organization.
\newblock {\em Frontiers in systems neuroscience}, 5:80.

\bibitem[Kostadinov et~al., 2019]{kostadinov2019predictive}
Kostadinov, D., Beau, M., Pozo, M.~B., and H{\"a}usser, M. (2019).
\newblock Predictive and reactive reward signals conveyed by climbing fiber
  inputs to cerebellar purkinje cells.
\newblock {\em Nature Neuroscience}, 22(6):950--962.

\bibitem[Linsker, 1988]{linsker1988self}
Linsker, R. (1988).
\newblock Self-organization in a perceptual network.
\newblock {\em Computer}, 21(3):105--117.

\bibitem[Maass, 2014]{maass2014noise}
Maass, W. (2014).
\newblock Noise as a resource for computation and learning in networks of
  spiking neurons.
\newblock {\em Proceedings of the IEEE}, 102(5):860--880.

\bibitem[Markram and Tsodyks, 1996]{markram1996redistribution}
Markram, H. and Tsodyks, M. (1996).
\newblock Redistribution of synaptic efficacy between neocortical pyramidal
  neurons.
\newblock {\em Nature}, 382(6594):807--810.

\bibitem[Mayr et~al., 2019]{mayr2019spinnaker}
Mayr, C., Hoeppner, S., and Furber, S. (2019).
\newblock Spinnaker 2: A 10 million core processor system for brain simulation
  and machine learning.
\newblock {\em arXiv preprint arXiv:1911.02385}.

\bibitem[Millidge et~al., 2020]{millidge2020predictive}
Millidge, B., Tschantz, A., and Buckley, C.~L. (2020).
\newblock Predictive coding approximates backprop along arbitrary computation
  graphs.
\newblock {\em arXiv preprint arXiv:2006.04182}.

\bibitem[Mnih and Gregor, 2014]{MnihGregor:14}
Mnih, A. and Gregor, K. (2014).
\newblock Neural variational inference and learning in belief networks.
\newblock {\em arXiv preprint arXiv:1402.0030}.

\bibitem[Monday et~al., 2018]{monday2018long}
Monday, H.~R., Younts, T.~J., and Castillo, P.~E. (2018).
\newblock Long-term plasticity of neurotransmitter release: emerging mechanisms
  and contributions to brain function and disease.
\newblock {\em Annual review of neuroscience}, 41:299--322.

\bibitem[Neftci et~al., 2016]{neftci2016stochastic}
Neftci, E.~O., Pedroni, B.~U., Joshi, S., Al-Shedivat, M., and Cauwenberghs, G.
  (2016).
\newblock Stochastic synapses enable efficient brain-inspired learning
  machines.
\newblock {\em Frontiers in neuroscience}, 10:241.

\bibitem[Nessler et~al., 2013]{NesslerETAL:13}
Nessler, B., Pfeiffer, M., Buesing, L., and Maass, W. (2013).
\newblock Bayesian computation emerges in generic cortical microcircuits
  through spike-timing-dependent plasticity.
\newblock {\em PLoS Computational Biology}, 9(4):e1003037.

\bibitem[Nessler et~al., 2009]{NesslerETAL:08}
Nessler, B., Pfeiffer, M., and Maass, W. (2009).
\newblock Hebbian learning of bayes optimal decisions.
\newblock {\em Adv. Neura.l Inf. Process. Syst.}, 21:1169--1176.

\bibitem[Oertner et~al., 2002]{oertner2002facilitation}
Oertner, T.~G., Sabatini, B.~L., Nimchinsky, E.~A., and Svoboda, K. (2002).
\newblock Facilitation at single synapses probed with optical quantal analysis.
\newblock {\em Nature neuroscience}, 5(7):657--664.

\bibitem[Pecevski et~al., 2014]{pecevski2014nevesim}
Pecevski, D., Kappel, D., and Jonke, Z. (2014).
\newblock {NEVESIM}: Event-driven neural simulation framework with a python
  interface.
\newblock {\em Frontiers in Neuroinformatics}, 8:70.

\bibitem[Pecevski and Maass, 2016]{pecevski2016learning}
Pecevski, D. and Maass, W. (2016).
\newblock Learning probabilistic inference through spike-timing-dependent
  plasticity.
\newblock {\em eNeuro}, 3(2).

\bibitem[Peyser et~al., 2017]{peyser2017nest}
Peyser, A., Deepu, R., Mitchell, J., Appukuttan, S., Schumann, T., Eppler,
  J.~M., Kappel, D., Hahne, J., Zajzon, B., Kitayama, I., et~al. (2017).
\newblock Nest 2.14. 0.
\newblock Technical report, J{\"u}lich Supercomputing Center.

\bibitem[Pfister and Gerstner, 2006a]{pfister2006beyond}
Pfister, J.-P. and Gerstner, W. (2006a).
\newblock Beyond pair-based stdp: A phenomenological rule for spike triplet and
  frequency effects.
\newblock In {\em Advances in neural information processing systems}, pages
  1081--1088.

\bibitem[Pfister and Gerstner, 2006b]{pfister2006triplets}
Pfister, J.-P. and Gerstner, W. (2006b).
\newblock Triplets of spikes in a model of spike timing-dependent plasticity.
\newblock {\em Journal of Neuroscience}, 26(38):9673--9682.

\bibitem[Ramstead et~al., 2016]{ramstead2016cultural}
Ramstead, M.~J., Veissi{\`e}re, S.~P., and Kirmayer, L.~J. (2016).
\newblock Cultural affordances: Scaffolding local worlds through shared
  intentionality and regimes of attention.
\newblock {\em Frontiers in psychology}, 7:1090.

\bibitem[Ramstead et~al., 2018]{ramstead2018answering}
Ramstead, M. J.~D., Badcock, P.~B., and Friston, K.~J. (2018).
\newblock Answering schr{\"o}dinger's question: A free-energy formulation.
\newblock {\em Physics of life reviews}, 24:1--16.

\bibitem[Rao and Ballard, 1999]{rao1999predictive}
Rao, R.~P. and Ballard, D.~H. (1999).
\newblock Predictive coding in the visual cortex: a functional interpretation
  of some extra-classical receptive-field effects.
\newblock {\em Nature neuroscience}, 2(1):79.

\bibitem[Rezende and Gerstner, 2014]{RezendeGerstner:14}
Rezende, D.~J. and Gerstner, W. (2014).
\newblock Stochastic variational learning in recurrent spiking networks.
\newblock {\em Frontiers in Computational Neuroscience}, 8:38.

\bibitem[Rusakov et~al., 2020]{rusakov2020noisy}
Rusakov, D.~A., Savtchenko, L.~P., and Latham, P.~E. (2020).
\newblock Noisy synaptic conductance: bug or a feature?
\newblock {\em Trends in Neurosciences}.

\bibitem[Szavits-Nossan and Evans, 2015]{szavits2015inequivalence}
Szavits-Nossan, J. and Evans, M.~R. (2015).
\newblock Inequivalence of nonequilibrium path ensembles: the example of
  stochastic bridges.
\newblock {\em Journal of Statistical Mechanics: Theory and Experiment},
  2015(12):P12008.

\bibitem[Tishby et~al., 2000]{tishby2000information}
Tishby, N., Pereira, F.~C., and Bialek, W. (2000).
\newblock The information bottleneck method.
\newblock {\em arXiv preprint physics/0004057}.

\bibitem[Toyoizumi et~al., 2005]{toyoizumi2005generalized}
Toyoizumi, T., Pfister, J.-P., Aihara, K., and Gerstner, W. (2005).
\newblock Generalized bienenstock--cooper--munro rule for spiking neurons that
  maximizes information transmission.
\newblock {\em Proceedings of the National Academy of Sciences},
  102(14):5239--5244.

\bibitem[Urbanczik and Senn, 2014]{urbanczik2014learning}
Urbanczik, R. and Senn, W. (2014).
\newblock Learning by the dendritic prediction of somatic spiking.
\newblock {\em Neuron}, 81(3):521--528.

\bibitem[Whittington and Bogacz, 2017]{whittington2017approximation}
Whittington, J.~C. and Bogacz, R. (2017).
\newblock An approximation of the error backpropagation algorithm in a
  predictive coding network with local hebbian synaptic plasticity.
\newblock {\em Neural computation}, 29(5):1229--1262.

\bibitem[Yang and Xu-Friedman, 2013]{yang2013stochastic}
Yang, H. and Xu-Friedman, M.~A. (2013).
\newblock Stochastic properties of neurotransmitter release expand the dynamic
  range of synapses.
\newblock {\em Journal of Neuroscience}, 33(36):14406--14416.

\bibitem[Yang and Calakos, 2013]{yang2013presynaptic}
Yang, Y. and Calakos, N. (2013).
\newblock Presynaptic long-term plasticity.
\newblock {\em Frontiers in synaptic neuroscience}, 5:8.

\end{thebibliography}
\end{document}